\documentclass[a4paper,fleqn,usenatbib]{mnras}

\usepackage{newtxtext,newtxmath}

\usepackage[T1]{fontenc}
\usepackage{ae,aecompl}

\usepackage{graphicx}	
\usepackage{amsmath}	
\usepackage{amssymb}	

\usepackage{color}

\def\mnras{MNRAS}
\def\aj{AJ}
\def\aap{A\&A}
\def\apj{ApJ}
\def\apjl{ApJ}
\def\apjs{ApJS}
\def\araa{ARA\&A}
\def\pasp{PASP}
\def\nat{Nature}

\newcommand{\angstrom}{\mbox{\normalfont\AA}}

\newcommand{\msun}{\mbox{M$_{\sun}$ }}

\newcommand{\lsunend}{\mbox{L$_{\sun}$}}
\newcommand{\lir}{\mbox{L$_{\rm IR}$}}

\newcommand{\htwo}{\mbox{H$_2$}}

\newcommand{\z}{\mbox{$z$}}

\newcommand{\zsim}{\mbox{$z\sim$ }}

\newcommand{\sunrise}{\mbox{\sc sunrise}}

\newcommand{\grasil}{\mbox{\sc grasil}}

\newcommand{\hyperion}{\mbox{\sc hyperion}}
\newcommand{\pd}{\mbox{\sc powderday}}

\newcommand{\gadget}{\mbox{\sc gadget-3}}

\newcommand{\galform}{\mbox{\sc galform}}
\newcommand{\gizmo}{\mbox{\sc gizmo}}

\newcommand{\music}{\mbox{\sc music}}
\newcommand{\caesar}{\mbox{\sc caesar}}
\newcommand{\mufasa}{\mbox{\sc mufasa}}
\newcommand{\fire}{\mbox{\sc fire}}

\newcommand{\fsps}{\mbox{\sc fsps}}
\newcommand{\pyfsps}{\mbox{\sc python-fsps}}

\newcommand{\yt}{\mbox{\sc yt}}

\newcommand{\irxb}{\mbox{IRX-$\beta$}}

\title[The \irxb \ Relation in Galaxies]{The IRX-$\beta$ Dust
  Attenuation Relation in Cosmological Galaxy Formation Simulations}

\author[D. Narayanan et al.]{Desika Narayanan$^{1}$\thanks{E-mail:
    desika.narayanan@ufl.edu}, Romeel Dav\'e$^{2,3,4,5}$, Benjamin
  Johnson$^{6}$, Robert Thompson$^{7}$, \newauthor Charlie
  Conroy$^{6}$, \& James E. Geach$^{8}$\\$^{1}$Department of Astronomy,
  University of Florida, 211 Bryant Space Science Center, Gainesville,
  FL, 32611, USA\\$^{2}$University of the Western Cape, Bellville,
  Cape Town 7535, South Africa\\${^3}$South African Astronomical
  Observatories, Observatory, Cape Town 7925, South
  Africa\\$^4$African Institute for Mathematical Sciences, Muizenberg,
  Cape Town 7945, South Africa\\$^5$Center for Computational
  Astrophysics, Simons Foundation, New York, New
  York\\${^6}$Harvard-Smithsonian Center for Astrophysics, 60 Garden
  Street, Cambridge, MA 02138 \\${^7}$Portalarium, 3410 Far West Blvd,
  Austin, TX 78731\\${^8}$Centre for Astrophysics Research, School of
  Physics, Astronomy \& Mathematics, University of Hertfordshire,
  Hatfield, AL10 9AB UK}%

\begin{document}

\date{Submitted to MNRAS}

\pagerange{\pageref{firstpage}--\pageref{lastpage}} \pubyear{2015}

\maketitle

\label{firstpage}

\begin{abstract}
We utilise a series of high-resolution cosmological zoom simulations
of galaxy formation to investigate the relationship between the
ultraviolet (UV) slope, $\beta$, and the ratio of the infrared
luminosity to UV luminosity (IRX) in the spectral energy distributions
(SEDs) of galaxies.  We employ dust radiative transfer calculations in
which the SEDs of the stars in galaxies propagate through the dusty
interstellar medium.  Our main goals are to understand the origin of,
and scatter in the IRX-$\beta$ relation; to assess the efficacy of
simplified stellar population synthesis screen models in capturing the
essential physics in the \irxb \ relation; and to understand
systematic deviations from the canonical local \irxb \ relations in
particular populations of high-redshift galaxies.  Our main results
follow.  Galaxies that have young stellar populations with relatively
cospatial UV and IR emitting regions and a Milky Way-like extinction
curve fall on or near the standard Meurer relation. This behaviour is
well captured by simplified screen models.  Scatter in the \irxb
\ relation is dominated by three major effects: (i) older stellar
populations drive galaxies below the relations defined for local
starbursts due to a reddening of their intrinsic UV SEDs; (ii) complex
geometries in high-\z \ heavily star forming galaxies drive galaxies
toward blue UV slopes owing to optically thin UV sightlines; (iii)
shallow extinction curves drive galaxies downward in the \irxb \ plane
due to lowered NUV/FUV extinction ratios. We use these features of the
UV slopes of galaxies to derive a fitting relation that reasonably
collapses the scatter back toward the canonical local
relation. Finally, we use these results to develop an understanding
for the location of two particularly enigmatic populations of galaxies
in the \irxb \ plane: $\zsim 2-4$ dusty star forming galaxies, and
$z>5$ star forming galaxies.

\end{abstract}

\section{Introduction}
Measuring the evolution of the cosmic star formation rate density
remains both a challenge and fundamental goal of galaxy evolution
astrophysics \citep[see ][for a recent review]{madau14a}.  Essentially
all tracers of a galaxy's star formation rate (SFR) measure the
luminosity (either directly or reprocessed) from massive, short-lived
stars.  Because the ultraviolet (UV) continuum from actively
star-forming galaxies is dominated by massive stars, direct
measurements of UV starlight from galaxies (alongside an assumption of
a stellar initial mass function) serves as one of the more straight
forward methods for deriving a galaxy's SFR
\citep[e.g.][]{kennicutt12a}.  This said, obscuring dust in the
interstellar medium of galaxies preferentially absorbs UV radiation,
and therefore complicates the interpretation of SFRs calibrated from
UV measurements.

The observed relationship between the rest-frame UV continuum slope
from galaxies ($\beta$, where $f_\lambda \propto \lambda^\beta$), and
the ratio of the infrared luminosity to the UV luminosity (IRX) has
been commonly employed as a tool for accounting for the dust
obscuration of UV and optical radiation.  In seminal work,
\citet{calzetti97a} and \citet{meurer99a} found that local starburst
galaxies lie on a well-defined sequence in the \irxb \ plane.  In
principle, this notional relationship is quite powerful: even when
infrared data is unavailable to constrain the true transfer of energy
from UV photons to thermal dust continuum, an \irxb \ relationship can
clarify the degree of dust obscuration from observations where only UV
measurements are available.

Indeed, a large number of studies have employed this technique to
infer the dust content of $\zsim 2-6$ galaxies for a range of
applications.  Amongst many others, these include: constraining the
cosmic star formation rate density \citep{bouwens14a,bouwens15a},
cosmic star formation rate functions
\citep[e.g.][]{bothwell11a,smit12a,smit15a}, the evolution of specific
star formation rates \citep[e.g.][]{gonzalez10a,gonzalez14a}, the
growth of galaxy dust content \citep{finkelstein12a}, physical
properties of individual galaxy populations
\citep[e.g.][]{stark09a,ho10a,williams14a}, and SFR tracers themselves
\citep[e.g.][]{treyer07a,kennicutt09a,hao11a}.  The rationale for
using the \irxb \ relation across a diverse range of observations has
been, in part, bolstered by a number of observations that show the
relation holding for samples of galaxies between \zsim 2-4
\citep[e.g.][]{reddy08a,reddy12a,pannella09a,seibert02a,heinis13a,to14a,bourne16a}.

At the same time, a number of observations have called into
question the relationship between $\beta$ and the infrared excess.
For example, at low-redshift, \citet{goldader02a} and
\citet{howell10a} find that  luminous infrared galaxies (LIRGs; \lir $>
10^{11}$ \lsunend) and ultraluminous infrared galaxies (ULIRGs; \lir
$>10^{12}$ \lsunend) are offset from the canonical
\irxb \ relations, with either an excess of IRX at a fixed $\beta$, or
bluer UV spectral slopes at a fixed IRX (i.e. they lie above the
relationship).  

This offset extends to high-redshift dusty galaxies as well.  For
example, while \citet{bourne16a} found that lower luminosity galaxies
at \zsim 2 present similarly as the \citet{meurer99a} sample in the
\irxb \ plane, higher luminosity dusty star forming galaxies tend to
exhibit bluer UV continuum slopes than the canonical relationship at a
given IRX value.  A similar trend is observed by \citet{reddy10a},
\citet{penner12a}, \citet{oteo13a, watson15a}, and \citet{casey14b} in high
redshift ($z \ga 2$) dusty star forming galaxies.  \citet{casey14b}
explicitly demonstrate that galaxies of increasing infrared luminosity
at $z \ga 0.5$ have increasingly blue UV SEDs.  We will return to this
point later in this paper.

Offsets from the \citet{meurer99a} local relation are not exclusive to
dusty star forming galaxies.  Metal-poor systems such as the Small and
Large Magallenic Clouds (SMC and LMC respectively) have redder colours
than more metal rich galaxies on the \irxb \ plane
\citep[e.g.][]{bell02a,buat05a}.  Similarly, some high-$z$ Lyman Break
Galaxies are more consistent with the SMC/LMC \irxb \ curve than the
\citet{meurer99a}-defined starburst locus
\citep[e.g.][]{reddy10a,koprowski16a}.  Even some so-called ``normal''
galaxies at high-redshift (those not currently undergoing a starburst event, with
$\sim$solar metallicity) have redder UV continuum slopes than the
canonical Meurer relation, and lie in between the Meurer relation and
SMC/LMC curves \citep[e.g.][]{buat05a,seibert05a,boquien09a,casey14b,pope17a}.

With the advent of the Atacama Large Millimetre Array (ALMA), dust
continuum detections (or sensitive upper limits) have now become
routine at the highest redshifts
\citep[e.g.][]{dunlop17a,rujopakarn16a}.  For example,
\citet{capak15a} and \citet{bouwens16a} employed $0.8-1.1$ mm
observations of $z \ga 4$ galaxies in order to infer their far
infrared luminosities (this requires an assumed dust temperature; we
will return to this point in \S~\ref{section:highz}).  From this data,
Capak et al. and Bouwens et al. find that these high-redshift galaxies
are systematically low in their dust content, lying below even the
SMC/LMC \irxb \ curves.

In this paper, we aim to develop a physical model for the origin of
and deviations from the \irxb \ relationship in galaxies.  To do this,
we will employ a series of high-resolution cosmological zoom galaxy
formation simulations that are coupled with both stellar population
synthesis and dust radiative transfer models.  This will allow us to
in effect ``observe'' the simulations, and therefore relate the
physical properties of the model galaxies to their observed UV
continuum and infrared luminosity properties.

We have three principle goals in this paper:
\begin{enumerate}
\item To understand the origin of and scatter in the \irxb \ relation
\item To asses the efficacy of simplified (i.e. screen) models in
  capturing the essential physics driving the \irxb \ relation
\item To utilise these models to understand systematic deviations from
  the canonical \irxb \ relations in particular populations of
  galaxies.
\end{enumerate}
To this end, we organize the paper as follows.  After describing our
numerical methodology and simulations (\S~\ref{section:methods}), we
deconstruct the \irxb \ relation, assessing the physical drivers of
dispersion in \S~\ref{section:fundamentals} and
\S~\ref{section:deconstruction}.  We then discuss individual galaxy
populations that appear to have systematic deviations from the \irxb
\ relation in \S~\ref{section:observations}.  Throughout these
sections, we will also include comparisons to simplified models in
which we place a dust screen in front of a simple stellar population
(hereafter ``screen models'').  In \S~\ref{section:discussion}, we
provide discussion, where we discuss a methodology to minimise the
uncertainty resultant from systematic deviations from the fiducial
\irxb \ relation, compare to other theoretical models, and assess the
$\tau_{\rm UV} - \beta$ and $\tau_{\rm UV}$-IRX relationships.
Finally, we summarise in \S~\ref{section:summary}

\section{Numerical Methods}
\label{section:methods}
\subsection{Overview}
Our overall aim is to simulate the rest-frame UV-millimetre wave SED of a
realistic sample of model galaxies in order to investigate the \irxb
\ relation in galaxies.  To do this, we model the formation and
evolution of galaxies utilising the cosmological ``zoom'' technique,
wherein individual galaxies are tracked in a cosmological simulation
at a higher resolution than the bulk of the cosmic volume.  This
technique, while computationally expensive, allows us to maintain a
relatively high mass resolution while still following the realistic
evolution of galaxies in their cosmic environment.

We couple these simulations with stellar population synthesis models
to model the spectral emission properties of the stars, and dust
radiative transfer simulations in order to calculate the dust
attenuation the stellar radiation encounters as it leaves the galaxy.
In the remainder of this section, we describe the details of these
galaxy formation simulations and dust radiative transfer calculations,
though note that the reader interested primarily in the main results
can skip this section without loss of continuity.

\subsection{Cosmological Zoom Galaxy Formation Simulations}
We conduct all of our galaxy formation simulations with a modified
version of the hydrodynamic code \gizmo \ \citep{hopkins14a}, which
builds off of much of the code base of \gadget \ \citep{springel05b}.
We simulate a $50 h^{-1}$ Mpc box from an initial redshift $z_{\rm
  init} = 249$ down to $z_{\rm final} = 0$, with initial conditions
generated with \music \citep{hahn11a}.  We employ a cosmology
$\Omega_\Lambda = 0.7$, $\Omega_{\rm b} = 0.048$, $H_0 = 68$ km
s$^{-1}$ Mpc$^{-1}$, and $\sigma_8 = 0.82$.  Our initial simulation
includes dark matter only, and included $512^3$ particles, resulting
in a dark matter mass resolution of $7.8 \times 10^8 h^{-1} M_\odot$.

We identify five arbitrary dark matter halos at redshift $z=z_{\rm
  final}$, where $z_{\rm final}$ is the final redshift of the zoom.
We have two categories of simulations: massive galaxies intended to
model a diverse range of physical properties at high redshift, as well
as a lower mass galaxy intended to represent 'normal' Milky Way like
galaxies in the present epoch.  The former category of simulations are
selected at (and eventually run to) $z_{\rm final} \approx 2$, while the
latter $z_{\rm final}=0$.  The massive simulations span more than a 
decade in mass, with $z=2$ halo masses of $M_{\rm DM} \approx 1 \times
10^{12}-3 \times 10^{13} \msun$, while the low-mass (low redshift)
simulation has a final halo mass comparable to the Milky Way's
\citep[e.g.][]{boylankolchin13a}.

Utilising \caesar \ \citep{thompson15b}, we follow the standard
procedure from \citet{hahn11a} in which we build an ellipsoidal mask
around all particles within $2.5 \times$ the radius of the maximum
distance (i.e. the farthest away) dark matter particle in the $z=z_{\rm final}$
halo, and define this as the Lagrangian high-resolution region to be
re-simulated at a higher-resolution.  When re-run to $z=z_{\rm final}$, we have
zero low-resolution particles contaminating our main halos within
three virial radii.

\begin{figure*}
  \includegraphics[scale=0.6,angle=270]{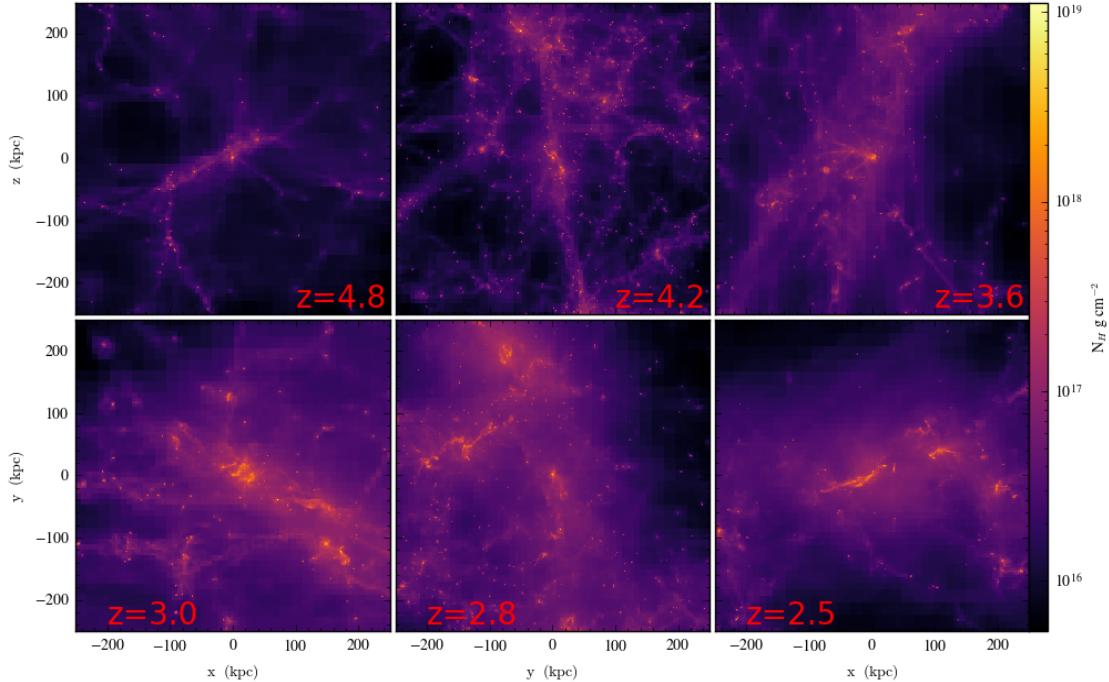}
  \vspace{-1.5cm}
\caption{ Gas surface density projections during the period $2 \la z
  \la 5$ of an arbitrary galaxy from our model sample (model mz5).
  The complex geometries will manifest themselves in the \irxb
  \ relation in \S~\ref{section:dsfgs}. \label{figure:morphology}}
\end{figure*}

\begin{table*}
\footnotesize
\caption{Summary of Model Galaxies.  $M_*$ and $M_{\rm halo}$ refer to
  the stellar and halo mass at $z=2$. }
\label{table:models}
\begin{tabular}{|c|c|c|c|c|}
\hline\hline
{\bf Model Name} &  {\bf $M_*$ (z=z$_{\rm final}$)}  & {\bf $M_{\rm halo}$ (z=z$_{\rm final}$)} & {\bf Final Redshift} &{\bf Notes} \\ 
                                  & $M_\odot$           & $M_\odot$                                   & &\\
\hline
\hline
mz0     & $1.7 \times 10^{11}$ & $2.95 \times 10^{13}$ &   2 & \\
mz0$_{\rm q}$  & $6 \times 10^{10}$ & $8.4 \times 10^{12}$ &   2.2 & Quenching model included\\
mz5     & $1.5 \times 10^{11}$ & $7.65 \times 10^{12}$ & 2 & \\
mz10    & $3.3 \times 10^{10}$ & $4.4 \times 10^{12}$ & 2& \\
mz45    & $1.2 \times 10^{10}$ & $9.95 \times 10^{11}$ & 2 & \\
z0mz401   & $2.6 \times 10^{10}$ & $2.05 \times 10^{12}$ & 0 &\\
\hline\hline
\end{tabular}
\end{table*}

We operate \gizmo \ in the meshless
finite mass (MFM) mode, which evolves the fluid in a mass-conserving
manner within individual mesh nodes.  We utilise a cubic spline kernel
with 64 neighbours in the MFM hydrodynamics, which is used to define
the volume partition between gas elements, and therefore the faces
over which the hydrodynamics is solved via the Riemann solver.
\citet{dave16a} have shown that relatively minimal differences exist
when using the MFM solvers in galaxy formation problems as a
pressure-entropy SPH \citep{hopkins13d}.  

Our simulations use the suite of physics described in the \mufasa
\ cosmological hydrodynamic simulations
\citep{dave16a,dave17a,dave17b}, and indeed our parent dark matter
simulation is generated with identical initial conditions as the $50$
Mpc $h^{-1}$ \mufasa \ simulation.  In this, stars form only in dense
molecular gas, where the \htwo \ fraction is dependent on the gas
metallicity and surface density following the \citet[][]{krumholz09b}
prescription \citep[e.g.][]{thompson14a}.  We impose a minimum
metallicity for star formation of $Z=10^{-3} Z_\odot$.

Star formation proceeds following a
\citet{schmidt59a} relation:
\begin{equation}
\frac{dM_*}{dt} = \epsilon_* \frac{\rho f_{\rm H2}}{t_{\rm dyn}}
\end{equation}
where $f_{\rm H2}$ is the mass fraction of gas in a given element that is
molecular, $t_{\rm dyn}$ is the local dynamical time scale, and
$\epsilon_*$ is the dimensionless star formation efficiency per free
fall time.  We set this latter quantity to be $\epsilon_* = 0.02$ as
motivated by observations and theoretical modeling
\citep{kennicutt98a,narayanan08b,narayanan12a,hopkins13b}. 

Feedback from young stars are modeled using a decoupled, two-phase
wind.  The wind physics are described in significant detail in
\citet{dave16a}, and we refer the reader to that work, summarising
only the most relevant details here.  The modeled stellar winds have a
probability for ejection which is modeled as a fraction ($\eta$) times
the star formation rate probability.  $\eta$ is modeled from the
best-fit relation by \citet{muratov15a}, from the Feedback in
Realistic Environments (\fire) simulation suite:
\begin{equation}
\eta = 3.55 \left(\frac{M_*}{10^{10}\msun}\right)^{-0.351}
\end{equation}
which is additionally motivated by analytic arguments
\citep{hayward17a}.  The stellar mass of the galaxy is determined
using a fast on-the-fly friends of friends finder \citep{dave16a}.
The ejection velocity scales in a modified way with the circular
velocity of the galaxy as well, following the scaling relations
derived by \citet{muratov15a}.  30\% of the winds are ejected hot,
with a thermal energy given by the difference between the supernova
energy and the kinetic energy of the launch. These winds are decoupled
from hydrodynamic forces or cooling until one of three conditions are
met: ($i$) its relative velocity compared to the surrounding gas is
less than half the local sound speed; ($ii$) the wind reaches a
density limit of $1\%$ of the critical density for star formation;
($iii$) $2\%$ of the Hubble time at launch has elapsed.

As we will discuss in \S~\ref{section:powderday}, the dust masses in
our radiative transfer simulations are tied to the metal content in
the ISM.  Feedback from longer-lived stars (Asymptotic Giant Branch
[AGB], and Type Ia supernovae) are included as well, following
\citet{bruzual03a} stellar evolution tracks with a \citet{chabrier03a}
initial mass function.  These delayed feedback sources deposit metals
into the interstellar medium.  We track the evolution of 11 elements:
H, He, C, N, O, Ne, Mg, Si, S, Ca and Fe.  The type II Supernovae
yields follow the \citet{nomoto06a} parameterisations as a function of
metallicity, though reduced by a factor $50\%$, owing to studies that
show these scalings result in metallicities a factor $\sim 2$ too high
as compared to the observed mass-metallicity relation \citep{dave11d}.  For Type Ia
supernovae, we employ the yields from \citet{iwamoto99a}, assuming 1.4
\msun \ of returned metals per SN1a event.  Chemical enrichment from
AGB stars follows the \citet{oppenheimer08a} lookup tables where the
yields are a function of age and metallicity.  

The default \mufasa \ model includes a heuristic quenching scheme in
which all gas that is not self-shielded is heated in massive halos
\citep[where the threshold mass for quenching is a function of
  redshift][]{dave16a}.  For the bulk of our models we do not include
this scheme as we endeavour to simulate a broad range of physical
conditions in star-forming galaxies, and including the quenching model
would limit the parameter space that we can study.  This said, in
order to build a substantial population of evolved stars (that will
enable us to understand the relationship between old stellar
populations and the \irxb \ relation), we include one model in which
this quenching model is included.

In Table~\ref{table:models}, we summarise the model zooms employed in
this study.  The dark matter particle masses are $M_{\rm DM} = 1
\times 10^{6} h^{-1} M_\odot$, and baryon particle masses $M_{\rm b} =
1.9 \times 10^{5} h^{-1} M_\odot$.  We employ adaptive gravitational
softening of all particles throughout the simulation
\citep{hopkins15a}, with minimum force softening lengths of $12, 3,$
and $3$ pc for dark matter, gas and stars respectively.  Each model
galaxy is evolved to $z=2$, outputting $85$ snapshots from $z \sim
30-2$.  In Figure~\ref{figure:morphology}, we show the gas surface
density projections for one of our model galaxies (mz5) as it evolves
from $z\sim 5-2$.

\begin{figure}
\includegraphics[scale=0.45]{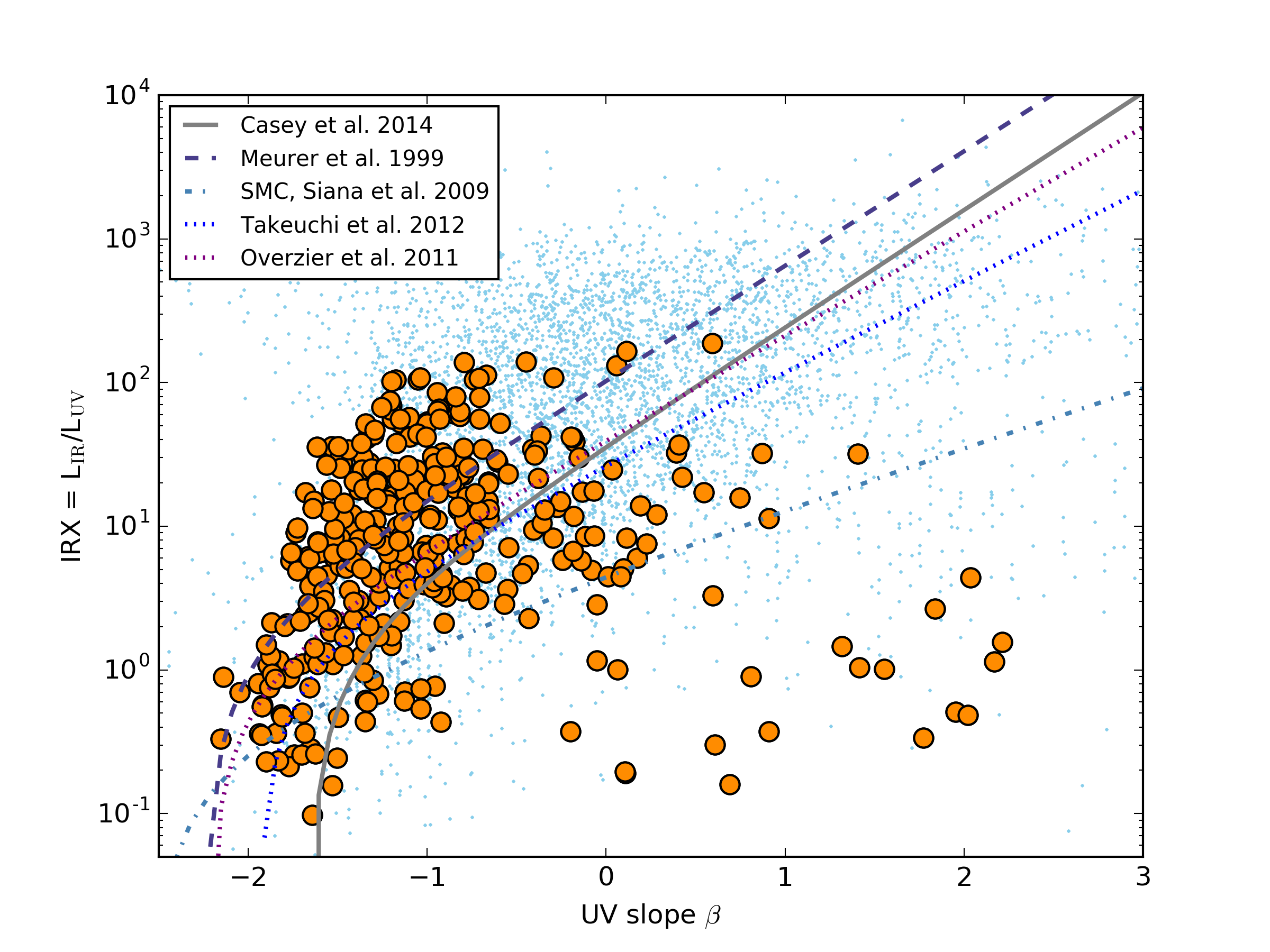}
\caption{\irxb \ relation for parent sample of model cosmological zoom
  galaxies.  The simulated galaxies are the large orange circles, and
  compared against the light blue points which are observations from
  $0\leq z \lesssim 6$
  \citep{casey14b,gildepaz07a,howell10a,reddy12a,penner12a,heinis13a,capak15a,bouwens16a}.
  The lines show the local reference relations (c.f. \S~\ref{section:reference}).\label{figure:irxbeta}}
\end{figure}

\begin{figure}
\includegraphics[scale=0.45]{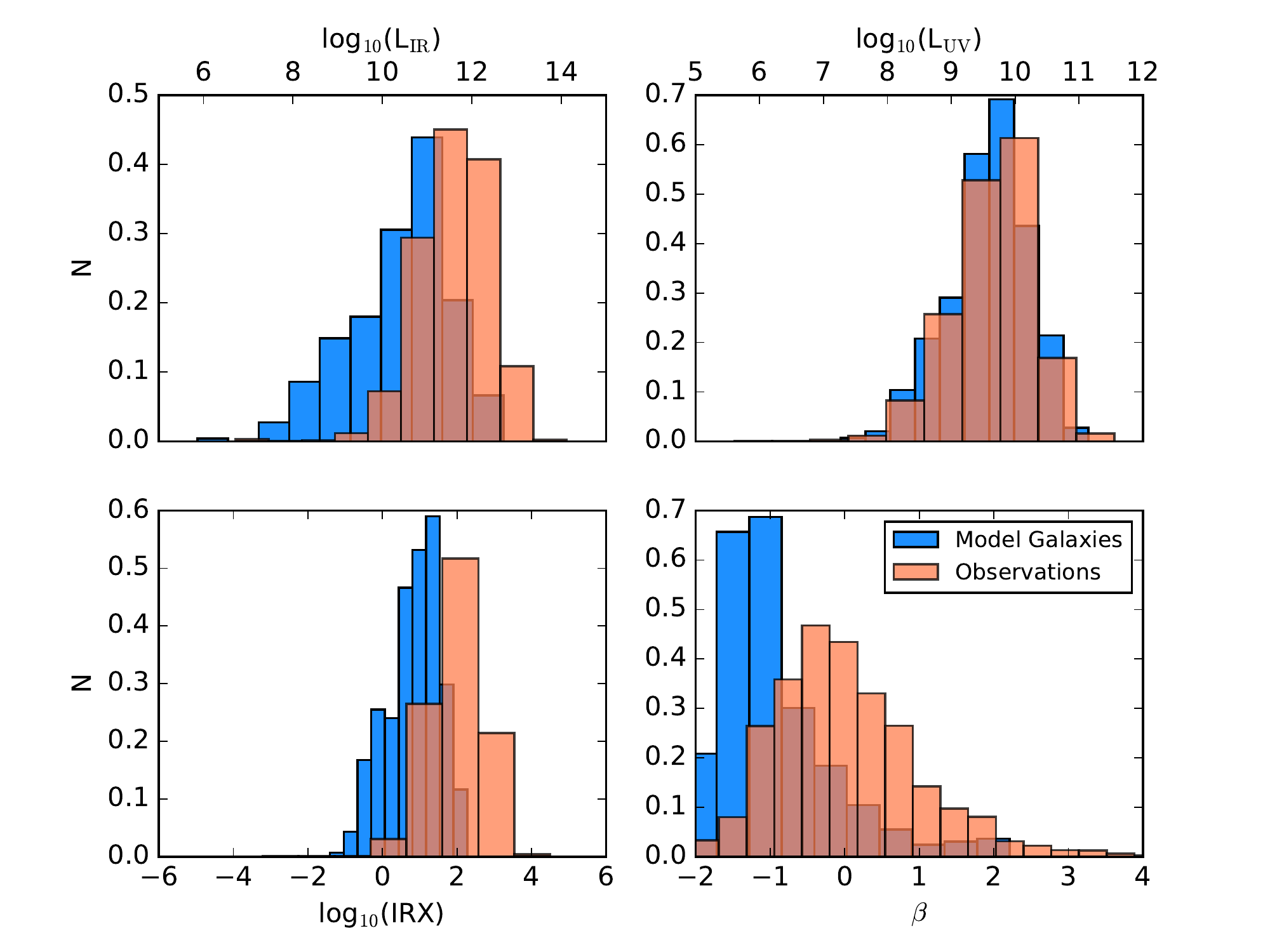}
\caption{Distributions of $L_{\rm IR}$, $L_{\rm UV}$, IRX and $\beta$
  for both sample of model galaxies (blue) compared to compilation of
  observations (orange).  Our modeled galaxies do not achieve as high
  a luminosity as the observed samples.  This results in lower peak
  IRX and $\beta$ values in our model galaxies compared to
  observations.  \label{figure:irxbeta_diagnostics}}
\end{figure}

\subsection{Dust Radiative Transfer}
\label{section:powderday}

With a galaxy sample in hand, we proceed to calculating the UV-far
infrared SED.  Our methodology contains two major steps: generating
the unattenuated stellar SEDs for all star particles in the
simulation, and then propagating these stellar SEDs through the dusty
interstellar medium to calculate the resultant attenuation.  This
process is performed on each snapshot at $z<10$ in post-processing.

These combined calculations are performed with \pd, an open-source
dust radiative transfer package first described in
\citet{narayanan15a} that wraps a number of publicly available codes
as described below. The stellar SEDs for each snapshot of each
modeled galaxy are calculated with the Flexible Stellar Population
Synthesis (\fsps) code \citep{conroy09b,conroy10a,conroy10b} via
\pyfsps\footnote{\url{https://github.com/dfm/python-fsps}}, a set of
Python hooks for \fsps.  The stellar SEDs are calculated as simple
stellar populations with their ages and metallicities taken directly
from the hydrodynamic simulation.  We assume a \citet{kroupa02a}
stellar initial mass function (IMF), and the Padova isochrones
\citep{marigo07a,marigo08a}.

We calculate the attenuation suffered by these stellar SEDs by
performing 3D Monte Carlo dust radiative transfer simulations.  The
metal mass from the galaxy formation simulations are projected onto an
adaptive grid with an octree memory structure.  The grid encompasses
$25$ kpc around the center of the galaxy.  Functionally, the octree is
constructed by placing the entire set of particles associated with the
halo within the radiative transfer simulation region into a single
cell, and then recursively subdividing this cell into octs until a
threshold maximum number of gas particles are in the cell (we employ
$n_{\rm subdivide, thresh} = 64$).  The physical properties of the gas
are projected using a spline smoothing kernel, and functionally done
with software within \yt \ \citep[][M.J. Turk et al. in
  prep.]{turk11a}.

The radiative transfer happens in a Monte Carlo fashion utilising the
dust radiative transfer code \hyperion \ \citep{robitaille11a}.
\hyperion \ uses the \citet{lucy99a} iterative algorithm for
determining the radiative equilibrium temperature in dust grains.  In
practice, photons are emitted from stellar sources, and absorbed,
scattered, and re-emitted by dust.  This process is iterated upon
until the dust temperature and radiation field are converged.  We
determine convergence when the energy absorbed by $99\%$ of the cells
has changed by less than $1\%$ between iterations.  Unless otherwise
specified (i.e. the tests done in \S~\ref{section:dust_composition}), we
assume a \citet{weingartner01a} grain size distribution, with $R
\equiv A_v/E(B-V) = 3.15$ (where $A_v$ is the visual extinction, and
$E(B-V)$ is the difference between the $B$ and $V$ band extinctions.
The emissivities are updated to include an approximation for
polycyclic aromatic hydrocarbons (PAHs) following
\citet{robitaille12a}.  The dust mass is assumed to be a constant
fraction (here, $40\%$) of the metal mass in each oct cell, where this
fraction is taken from constraints across a range of epochs
\citep{dwek98a,vladilo98a,watson11a}.

The result of the \pd \ calculations is a simulated SED for each
galaxy snapshot, spanning the wavelength range $912 \angstrom$ to $1$ mm.  It
is these SEDs that we use as both a comparison to observations, as
well as an interpretative tool for understanding the \irxb \ relation
in galaxies.  Example SEDs can be seen in Figure~\ref{figure:dtm}.

\section{\irxb \ Fundamentals}
\label{section:fundamentals}
\subsection{\irxb \ Calculations}
The infrared excess (IRX) is defined as:
\begin{equation}
{\rm IRX} \equiv \frac{L_{\rm IR}}{L_{\rm UV}}
\end{equation}
We define the infrared luminosity as the bolometric luminosity between
$8$ and $1000$ \micron, though note that our results are not
significantly changed if we use a minimum wavelength of $\lambda = 1
$\micron \ in our definition of $L_{\rm IR}$.  Here, we use the flux at
the closest wavelength output from our models to $1600 \angstrom$,
which is $1601.03 \angstrom$.

In order to calculate the UV continuum slope, $\beta$, we fit a simple
power law $f(\lambda) \propto \lambda^\beta$ between
$\lambda=1000-3000 \angstrom$ (where the flux is in units of erg
s$^{-1}$ cm$^{-2}$ $\angstrom^{-1}$).  We choose this relatively broad
range to marginalize the effects of the $2175 \angstrom$ dust feature,
when present.  

\subsection{The Reference Relations}
\label{section:reference}
Throughout the paper we will compare to a set of 'reference' relations
to help put our results into context.  These include: (i) the original
\citet{meurer99a} relation, which is a fit to observations of local
starburst galaxies; (ii) the \citet{casey14b} relation, which is
additionally a fit to low redshift galaxies, but includes a larger
dynamic range of SFRs as well as aperture-corrected data from
heterogeneous samples; (iii) recent calibrations by
\citet{takeuchi12a} and \citet{overzier11a}; and (iv) SMC curves
calculated by \citet{siana09a}.  For the SMC curves, Siana et
al. utilise constant star formation history \citet{bruzual03a}
population synthesis models ($t_{\rm age} = 300$ Myr) with an SMC
extinction curve.

\subsection{The Parent Sample in the \irxb \ Plane}
\label{section:parent_sample}
In \autoref{figure:irxbeta}, we show the location in the \irxb \ plane
of all snapshots of our simulated galaxies (orange points).
Alongside our simulated galaxies, we show a compilation of observed
data (blue), comprised of the data presented in \citet{casey14b},
which in itself is a compilation of new observations, and the samples
of \citet{gildepaz07a,howell10a,reddy12a,penner12a} and
\citet{heinis13a}.  We supplement these with more recent detections by
\citet{capak15a} and \citet{bouwens16a}.  We have omitted upper limits
for the time being, but will return to these later in this paper.  To
guide the eye, we additionally plot the reference
relations (c.f. \S~\ref{section:reference}).  The comparison of our
model galaxies to the observed \irxb \ plane presents a number of
salient points.

First, the simulated galaxies naturally lie on a locus similar to the
local reference relations, though there are a number of deviations
both at high IRX, and low IRX.  The dispersion is significant, and
owes to a range of physical effects that we will explore in turn in
this paper.

Second, while our galaxies follow the same generic trend as observed
galaxies, they do not fully cover the same extent in IRX and $\beta$
as observed galaxies.  We quantify this in
\autoref{figure:irxbeta_diagnostics}, where we show histograms of the
$L_{\rm IR}$, $L_{\rm UV}$, IRX and $\beta$ of both our model
galaxies, and the observed comparison samples.  While our galaxies are
comparably UV-luminous, we do not achieve infrared luminosities much
beyond $L_{\rm IR} \geq 10^{13} \lsunend$, and consequently do not
have model galaxies with as extreme IRX values as the most extreme
outliers in the observed sample.  Of course a lack of the dustiest
systems in our model sample also results in model galaxies restricted
to a a $\beta$ range $-2<\beta<1$, therefore falling short of the very
reddest UV-continuum slopes observed.

The lack of exact range matching is not a failure of the models, but
rather to be taken as a comparison of sample selections: our models
contain, on average, galaxies of roughly $5-10\times$ lower infrared
luminosity than the average observed galaxy that we compare to.  This
said, despite the fact that the simulations do not cover the entire
range of values seen in the observations, significant insight can
still be gained.

In the remainder of this paper, we will dissect
\autoref{figure:irxbeta}, aiming to understand both the origin of the
\irxb \ relation in galaxies, as well as trends that result in
deviations from the mean relation.  To do this, we will no longer
present the entire sample, but rather individual sub-samples
restricted by either physical or observable properties as it pertains
to the relevant discussion.

\section{Deconstructing the \irxb \ Relation}
\label{section:deconstruction}

\begin{figure*}
\includegraphics[scale=0.9]{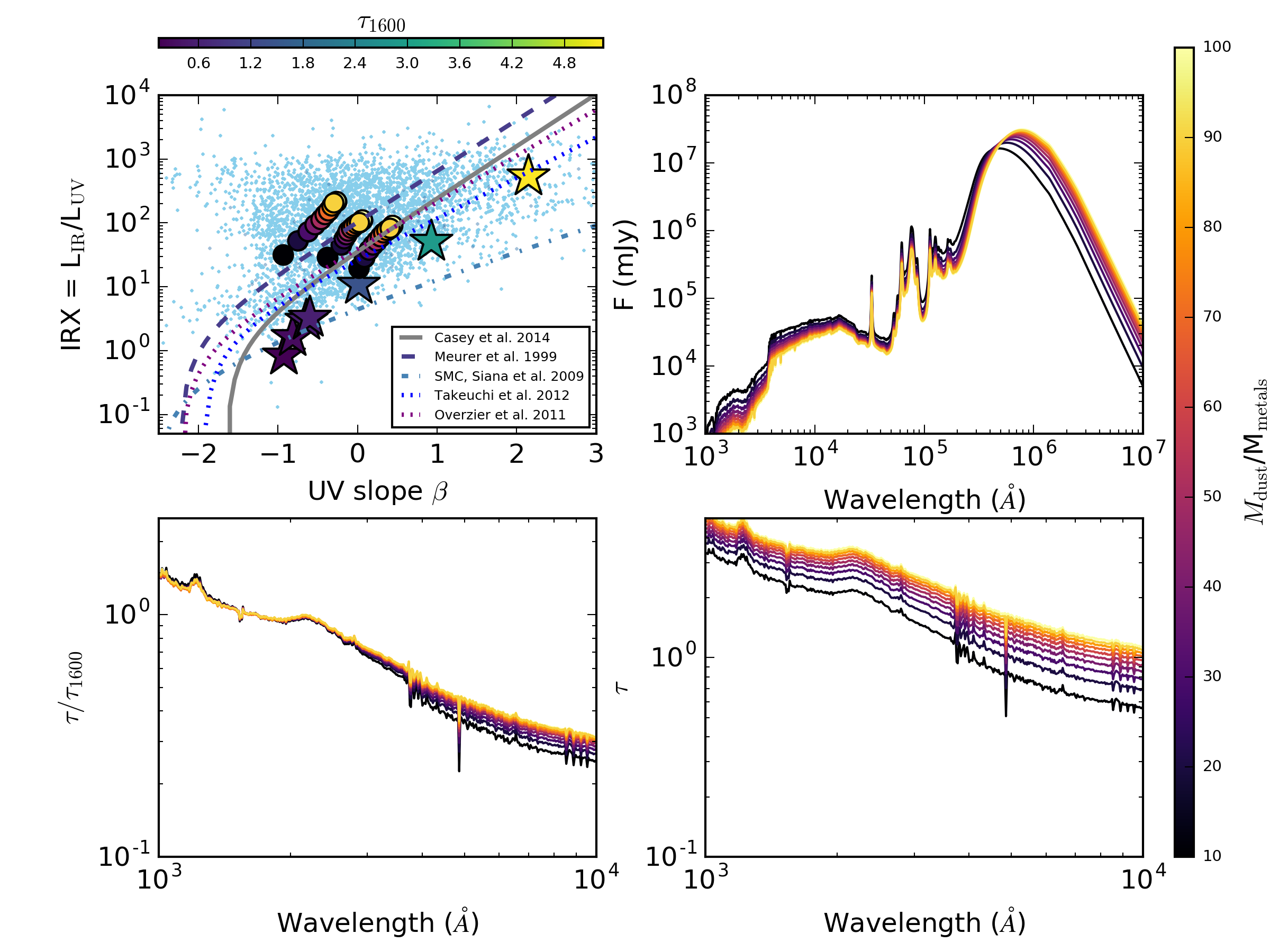}
\caption{Evolution of a model galaxy in the \irxb \ plane (and SED
  evolution) when manually increasing the dust mass of the system.
  {\bf Top Left:} Stars show \irxb \ relation for a simple stellar
  population with increasing dust optical depths (colourbar on top).
  Filled circles show three arbitrary simulated galaxies with
  increasing dust masses (colourbar on right).  {\bf Top Right:} UV-mm
  wave SED for just one of the three galaxies, with increasing dust
  content.  {\bf Bottom:} Optical depths both normalised by the $1600
  \angstrom$ optical depth (left) and not  (right).  \label{figure:dtm}
}
\end{figure*}

\begin{figure*}
  \includegraphics[]{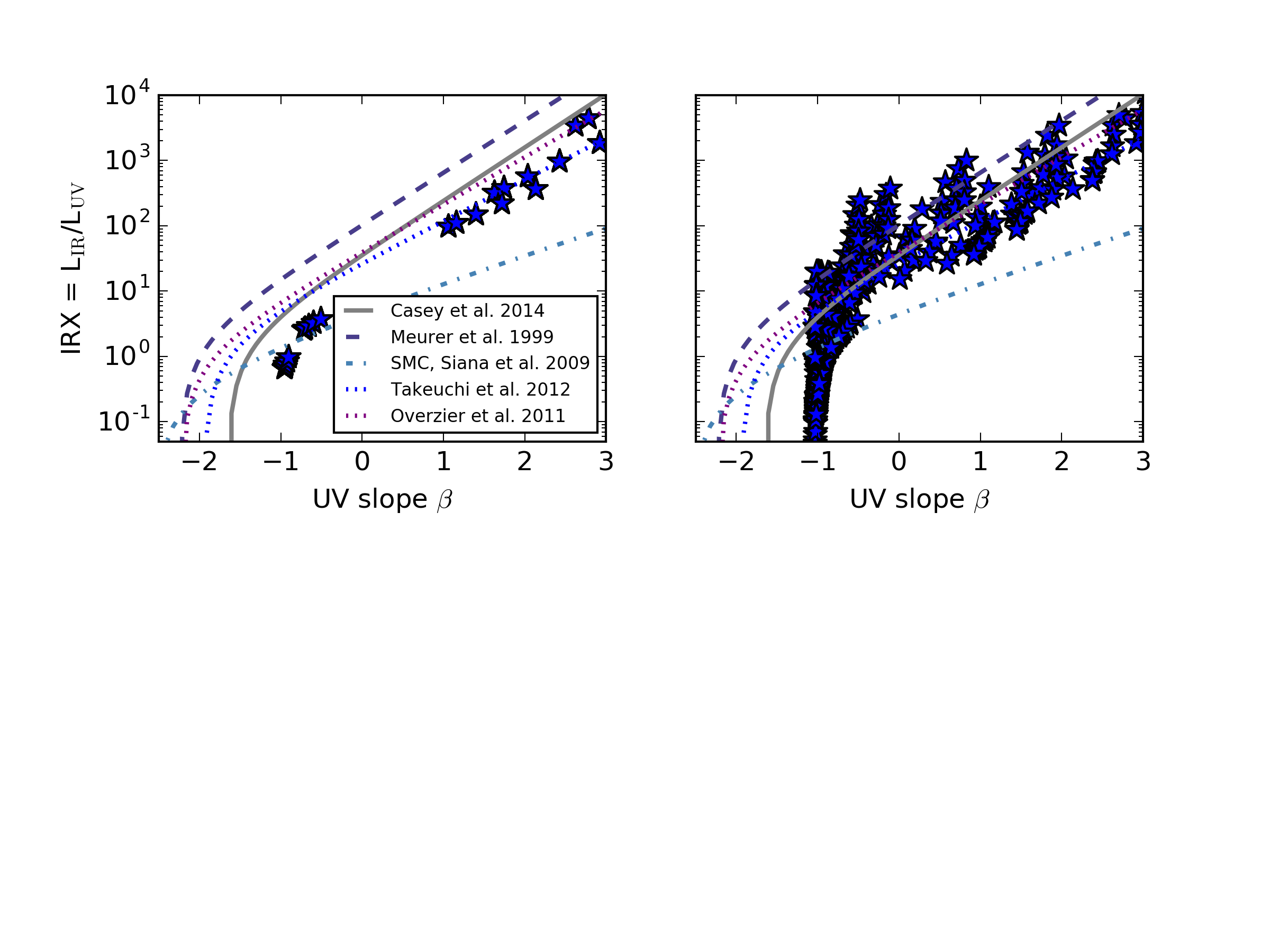}
  \vspace{-7cm}
  \caption{Numerical experiment conducted with \fsps \ in which we
    place a simple stellar population behind a uniform
    powerlaw dust screen with varying optical depths with
    the goal of understanding whether simple screen models can
    reproduce the dispersion in the observed \irxb \ relation. {\bf
      Left:} Here, we vary both the normalisation and power-law index
    of the dust law.  As the optical depth to UV photons rises, more
    power is transferred from the UV SED to the infrared, forcing the
    stellar population up the \irxb \ relation.  This said, the
    dispersion is relatively minimal in the resultant \irxb
    \ relation.  {\bf Right:} Same as left, but with a range of
    fractions of OB stars that escape their birth clouds, old stars
    that are not reddened by diffuse dust, dust attenuation powerlaw
    indices and optical depths.  Now, with complex star-dust
    geometries and attenuation laws, the \irxb \ normalisation shifts,
    causing significant scatter about the reference
    relations. \label{figure:fsps_full_irxb_range}}
\end{figure*}

\subsection{The Origin of the \irxb \ Relation}

We begin by understanding the origin of the reference \irxb
\ relations; we then proceed to deconstruct large deviations from
these relations.

The simplest place to start is with stellar population synthesis
experiments.  In the top left panel of Figure~\ref{figure:dtm}, we
have created a simple stellar population with \fsps \ that is covered
with a foreground dust screen and a stellar age $t_{\rm age} =
10^{-3}$ Gyr.  These are denoted with the star symbols (we will
discuss the circles, as well as the other panels in the plot shortly).
For maximal simplicity, we employ this dust screen to all stars,
and manually increase the normalisation (i.e. optical depth) of the
attenuation curve to study the location of the emergent SED in the
\irxb \ plane.  In our default population synthesis model shown here,
old stars ($t_{\rm age} > 10^7$ yr) do not see any dust,
although in this particular example we fix all stellar ages to much
less than this threshold value.  The dust attenuation curve is a
powerlaw with index $-0.7$ (i.e. $\tau(\lambda) \propto
\lambda^{-0.7}$), stellar initial mass function of a \citet{kroupa02a}
form, and solar metallicity.  The colours of the points denote the
monochromatic ($1600 \angstrom$) UV optical depth.  As the optical
depth for UV photons increases, the UV SED reddens, while at the same
time power is transferred from the UV to the IR.  This idealized
example serves as a control experiment in that it is unmuddied by the
complicating effects of a diverse stellar population, and the complex
star-dust geometry characteristic of real galaxies.

We further expand on this experiment by conducting a slightly more
complex numerical exploration utilising our model galaxies.  We now
turn to the filled circles in the top left panel of
Figure~\ref{figure:dtm}.  Here, we have selected three arbitrary
galaxies from our sample (all selected at the same redshift [$z \sim
  2.25$] from galaxies mz0, mz5 and mz10), and plot their location
using our fiducial parameters on the \irxb \ relation.  We then
manually decrease and increase the dust mass of these galaxies by
adjusting our dust to metals ratio and show the location of these
galaxies in the \irxb \ plane.  (We remind the reader that our
fiducial dust to metals ratio is $40\%$).  Alongside this, we show the
panchromatic SEDs for these points (top right), the UV optical depth
(bottom left; normalised by their $1600 \angstrom $ optical depth, and
the un-normalised optical depth (bottom right).

As the dust column increases in Figure~\ref{figure:dtm} the optical
depth seen by UV photons increases, and the UV SED flattens.  At the
same time, power is transferred from the UV into the infrared,
resulting in increased IRX values.   Of course,
Figure~\ref{figure:dtm}
represents contrived scenarios to show how galaxies may move along the
reference \irxb \ relations in the face of increasing dust column.
However, as is evident both from the observed data, as well as our
full suite of numerical simulations
(e.g. Figure~\ref{figure:irxbeta}), there is significant dispersion
about these reference relations.  In the remainder of this paper, we
explore the origin of these deviations from the reference \irxb
\ relations.

\subsection{Dispersion in the \irxb \ Relation}

\subsubsection{Can Simplified Screen Models Capture \irxb \ Dispersion?  The Role of Geometry}

Prior to understanding the origin of the dispersion in the \irxb
\ relation, it is first worth considering whether a simple screen
model at a single stellar age can capture a similar range of IRX and
$\beta$ values as both the cosmological zoom galaxy formation
simulations, and the observations.

In Figure~\ref{figure:fsps_full_irxb_range} we plot the results of
\fsps \ models akin to those presented in the top left panel of
Figure~\ref{figure:dtm}.  These models are a simple stellar population
with a uniform dust screen with a powerlaw attenuation curve. In the
left panel of Figure~\ref{figure:fsps_full_irxb_range}, we vary just
the dust attenuation powerlaw index (from [$-0.2,-1.2$]) alongside the
attenuation law normalisation.  While this model is able to span
arbitrarily large $\beta$ values, the very blue UV colours of high IRX
galaxies are not encapsulated in this model.  In other words, simple
screen geometries in population synthesis models fail to capture the
complex physics underlying the locations of galaxies on the \irxb
\ plane.

In the right panel of Figure~\ref{figure:fsps_full_irxb_range}, we
allow a fraction of young O and B stars ($t_{\rm age} < 10^7$ yr) in
the SPS models to run away from their birth clouds, as well as a
fraction of old ($t_{\rm age} > 10^7$ yr) stars to be decoupled from
the diffuse dust emission.  The goal here is to decouple the sources
of UV luminosity from the obscuring dust.  We allow both the fraction
of runaway OB stars and the fraction of decoupled old stars from dust
to vary between [$0,1$].  In this somewhat extreme experiment, it is
clear that the complex star-dust geometry is able to drive a large
range of \irxb \ values, and that the observed dispersion is due to a
decoupling of stars from the dust attenuation screen.

\begin{figure}
  \includegraphics[scale=0.45]{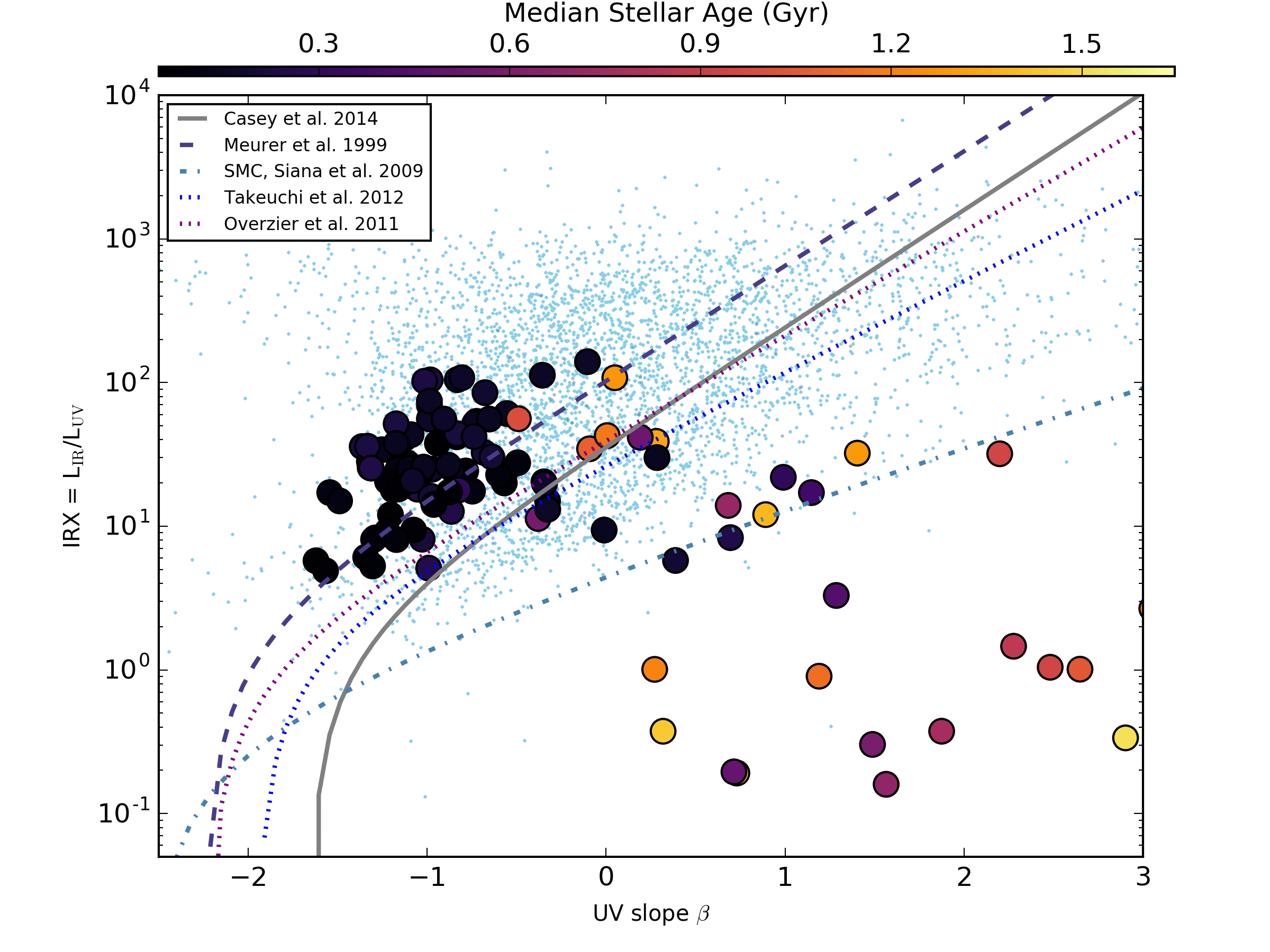}

  \caption{Impact of stellar age on location in the \irxb \ plane.
    Shown is a controlled experiment, comparing models mz0 and mz0\_q
    (i.e. identical models, though one includes a heuristic quenching
    model).  The points are colour-coded by their median stellar age.  Galaxies with
    older stellar populations lie below the local reference \irxb
    \ relations due to the shifting of the UV SED toward redder
    colours, and lack of young stars to dominate the UV SED
    (c.f. Figure~\ref{figure:fsps_stellarages}). \label{figure:simulations_stellarages}}
\end{figure}

\begin{figure*}
  \includegraphics[scale=0.8]{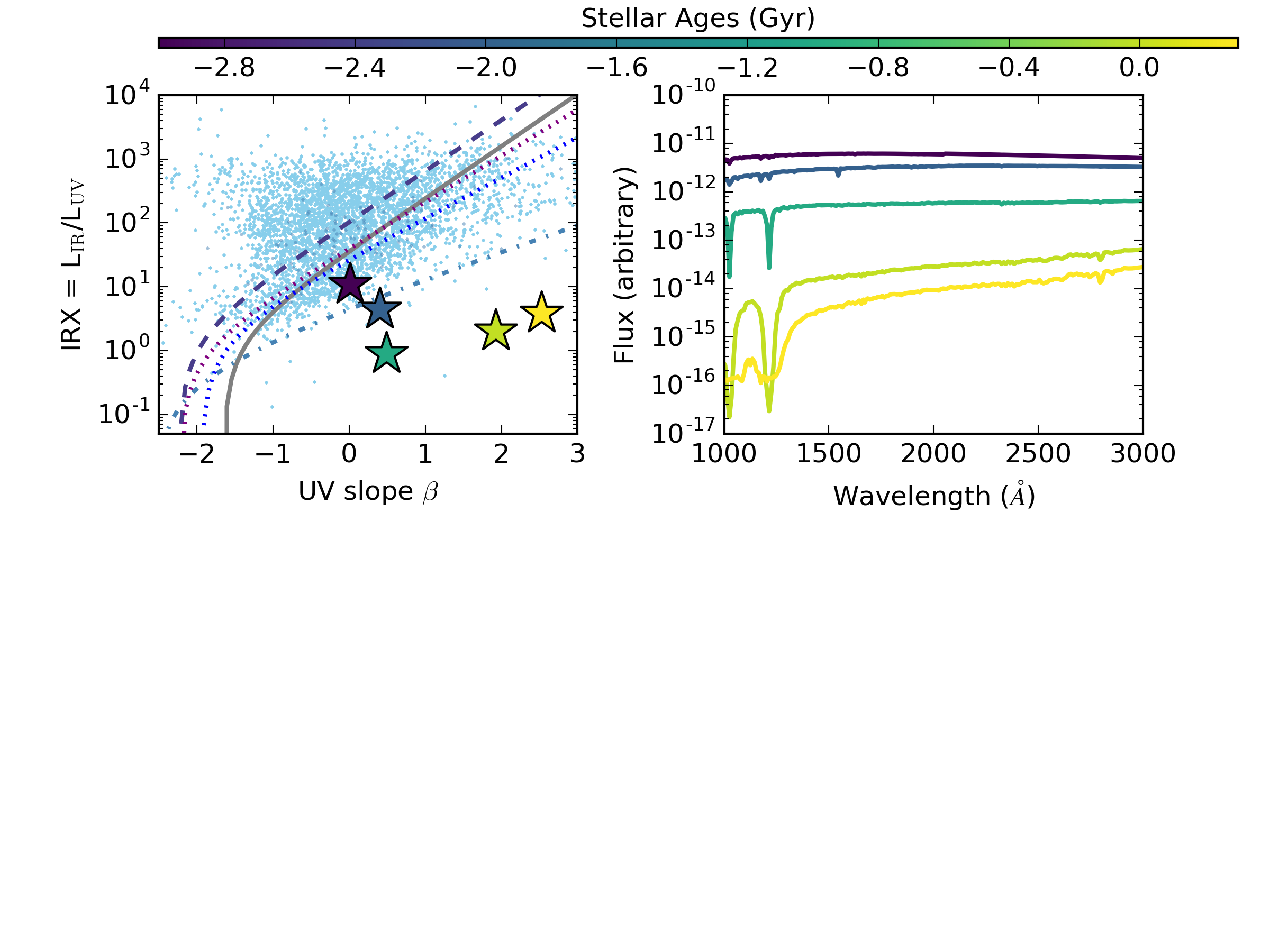}
\vspace{-5cm}
\caption{Results from a controlled \fsps \ numerical experiment in
  which we model the location of a simple stellar population behind a
  fixed dust screen as the stellar population ages.  The left panel
  shows the location of these stellar population on the \irxb \ plane
  (colour-coded by the stellar population age), while the right panel
  shows the UV SED for these populations.  As is evident, as the
  population ages, the UV SED reddens, pushing galaxies below the
  local reference \irxb \ relations.\label{figure:fsps_stellarages}}
\end{figure*}

\subsubsection{The Impact of Older Stellar Populations}
\label{section:oldstars}
We now return to Figure~\ref{figure:irxbeta}, and highlight the locus
of galaxies that lie below the reference \irxb \ relations.  These galaxies
predominantly arise from our halo0$_{\rm q}$ (quenched) model, and reflect
quenched galaxies with older stellar populations.  In general, older
stellar populations can drive red UV SEDs, even while maintaining
relatively low IRX values.  

To demonstrate this point more explicitly, in
Figure~\ref{figure:simulations_stellarages}, we show the results of a
controlled numerical experiment in which we compare the mz0 and mz0\_q
models.  The primary difference in these models is that the latter
includes our heuristic quenching model, and therefore has a
significant population of evolved stars.  The points are coloured by
their median stellar age.  As is clear, the galaxies with the oldest stellar
populations preferentially lie well below the reference \irxb
\ relations.

This effect owes to the reddening of UV SEDs in older stellar
populations.  To show this, in the left panel of
Figure~\ref{figure:fsps_stellarages}, we conduct a numerical
experiment with \fsps, where we evolve a simple stellar population
behind a dust screen.  The example is the same as in
Figure~\ref{figure:dtm}, though we fix the dust optical depth and vary
the stellar ages.  In the right panel of the same figure, we show the
UV and optical SEDs for this aging population.  As stellar populations
age, their UV SEDs shift toward redder $\beta$ slopes, as demonstrated
by the right panel of Figure~\ref{figure:fsps_stellarages}.  At the
same time, they are less likely to be obscured by dust (as a reminder,
in these \fsps \ calculations, stellar populations with $t_{\rm age} >
10^7$ yr are unobscured), and their IRX ratios drop.  (We note there
is a slight increase in IRX as the populations continue to age owing
to dramatic decreases in $L_{\rm UV}$).

Of course, using the median stellar age as a sole parameter for
describing deviations below the \irxb \ relation is a crude parameter.
What actually matters is the stellar population dominating the UV SED,
which is a function of the stellar age distribution.  Because of that,
there is significant dispersion in the ages of galaxies that lie below
the reference \irxb \ relations.  This said, the general point stands
that, on average, older stellar populations deviate toward redder UV
colours.  This point has been discussed by \citet{kong04a}, who
parameterised deviations from the \irxb \ relation in terms of a
stellar birthrate parameter, comparing the present SFR to the time
averaged past.

\subsubsection{Dust Composition and the $2175 \angstrom$ Bump}
\label{section:dust_composition}

A number of studies have underscored the importance of the underlying
extinction curve in driving how galaxies present in the \irxb \ plane.
For example, as previously discussed, \citet{siana09a} utilise
\citet{bruzual03a} calculations to demonstrate that an LMC and
SMC-like dust reddening curve result in \irxb \ relations that lie,
for the most part, below the standard \cite{meurer99a} relation.
Similarly, \citet{bell02a} demonstrate that sightlines in the LMC all
lie below the \citet{meurer99a} galaxies in the \irxb \ plane.

\begin{figure}
  \includegraphics[scale=0.45]{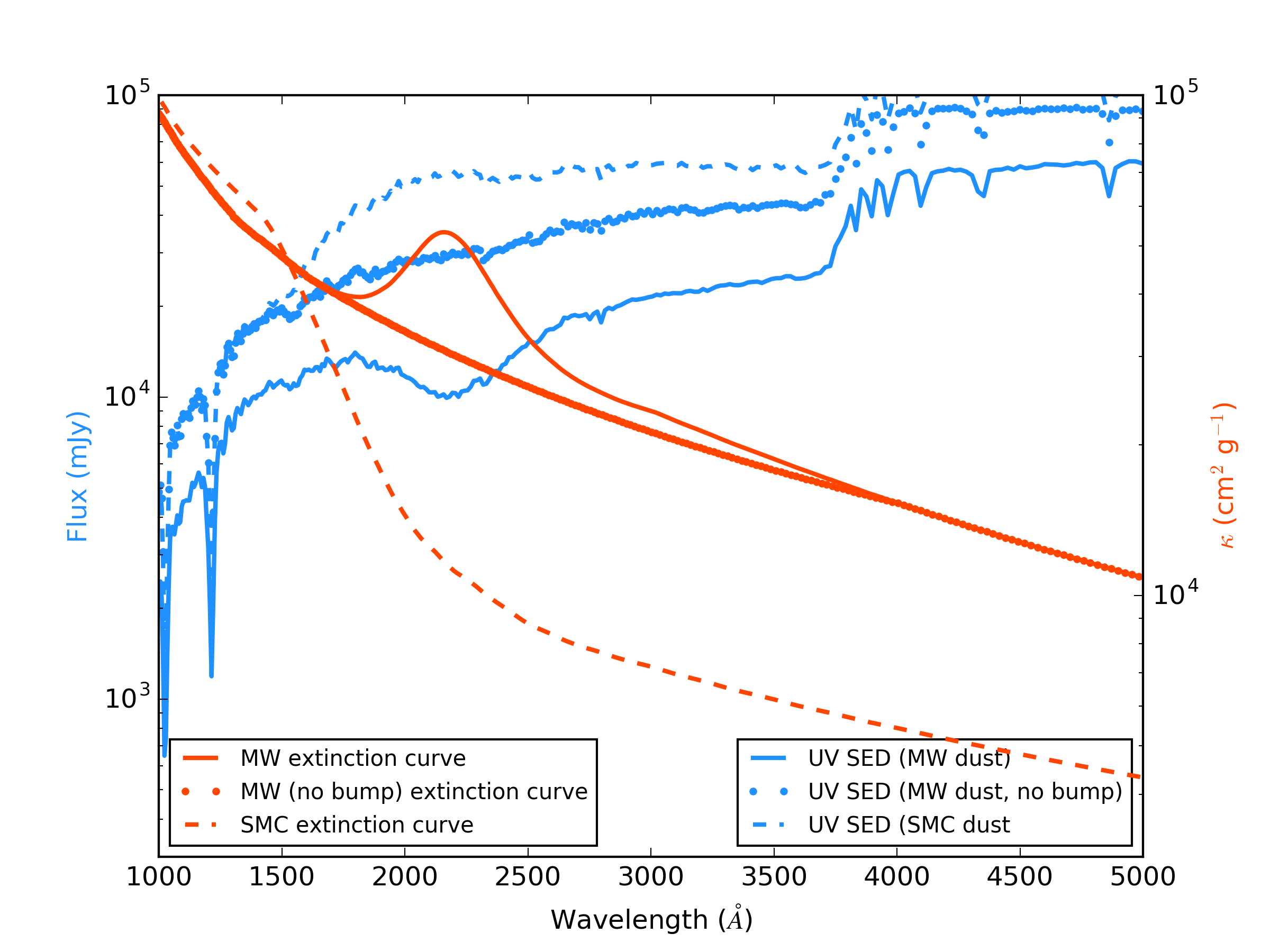}
  \caption{Tests varying the dust extinction curves in the models.
    Red lines (units on right axis) show the extinction curves the MW,
    MW with no dust bump, and the SMC.  Blue lines (units on left
    axis) show how the UV-optical SED for an example simulated galaxy
    changes when using these extinction curves.  Broadly, the MW (no
    bump) and SMC curves have less extinction in the NUV (as compared
    to the FUV) which allow more $2000-4000 \angstrom$ photons to
    escape the galaxy.  When fitting the UV SED over a range anchored
    at shorter wavelengths (e.g. $1000-3000 \angstrom$, this can cause
    the SED to appear redder, with steeper $\beta$
    slopes.  \label{figure:smc_uv_sed}}
\end{figure}

To understand how the intrinsic extinction curve impacts the observed
\irxb \ relation, we have conducted two numerical experiments on top
of our fiducial model (which, as a reminder, utilises a Milky Way
\citet{weingartner01a} $R_{\rm v} = 3.1$ extinction curve).  In our first
experiment, we utilise the same curve, though have eliminated the $2175
\angstrom$ \ extinction 'bump'. We remove the bump by linearly
interpolating the extinction curve between $1600-4000 \angstrom$ in
log-space.  This results in reduced attenuation in this wavelength
range.  For the second experiment, we utilise the
\citet{weingartner01a} SMC dust curve.  In
Figure~\ref{figure:smc_uv_sed}, we show these extinction curves (red
lines).

\begin{figure}
  \includegraphics[scale=0.45]{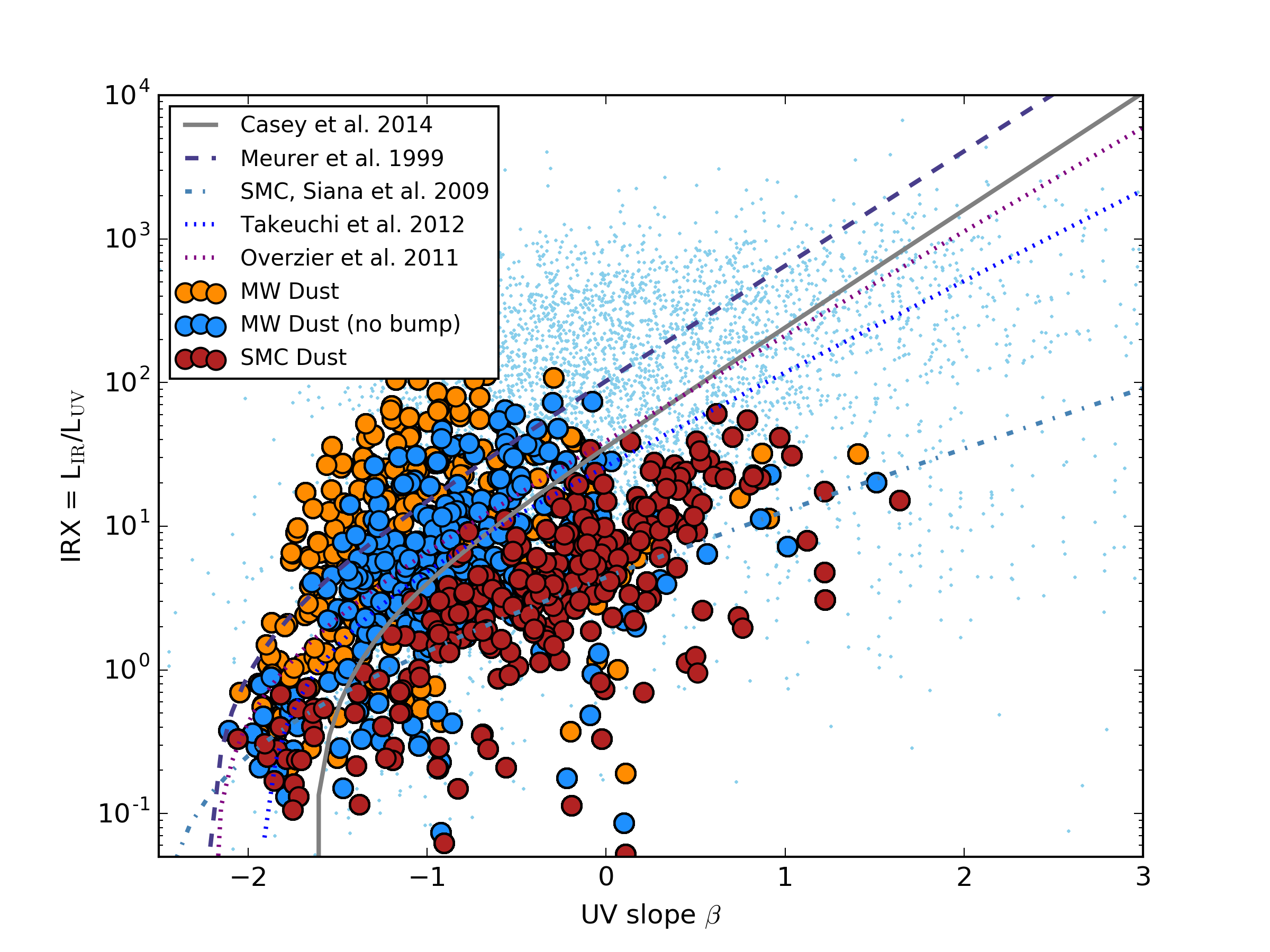}
  \caption{Impact of dust extinction curve on location in \irxb
    \ plane.  The orange points are our fiducial model (MW dust). The
    blue points are the same model, but using a MW extinction curve
    without a $2175 \angstrom$ UV ``bump''.  The red points show the
    impact of using an SMC extinction curve.  The reduction in UV
    extinction drives UV SEDs to lower IRX (owing to increased
    $L_{\rm UV}$), and redder $\beta$ (owing to more transparency in
    the $2000-4000 \angstrom$ window).  See text for details.
    \label{figure:smc}}
\end{figure}

In Figure~\ref{figure:smc}, we show the results from these models in
the \irxb \ plane.  Galaxies without a $2175 \angstrom$ bump (but
otherwise similar dust extinction properties to the Milky Way) has
both lower IRX and larger $\beta$ values than our fiducial model that
includes the UV extinction bump.  The SMC dust curve which has even
further reduced extinction in the UV presents with both even lower IRX
and redder $\beta$.

To understand the origin of the lower IRX and redder $\beta$ values
for the model without a UV bump and the SMC dust curve, we now turn to
the the UV SEDs (blue lines) presented in
Figure~\ref{figure:smc_uv_sed}.  At a fixed far ultraviolet (FUV)
opacity, the Milky Way dust curve with no bump has a lower near
ultraviolet (NUV) opacity, and the SMC curve a yet lower NUV opacity.
As this NUV extinction is reduced, the observed UV SED shows larger
powerlaw SED slopes owing to increased transparencies in the
NUV/optical bands.  Of course, the location of galaxies with the SMC dust curve is
degenerate with the impact of older stellar populations
(c.f. Figure~\ref{figure:fsps_stellarages}).  One must therefore
exercise caution in the interpretation of galaxies that lie below the
\irxb \ relation.

\section{Application to Observations}
\label{section:observations}

\subsection{Blue Dusty Star Forming Galaxies}
\label{section:dsfgs}

\begin{figure*}
\includegraphics[scale=0.9]{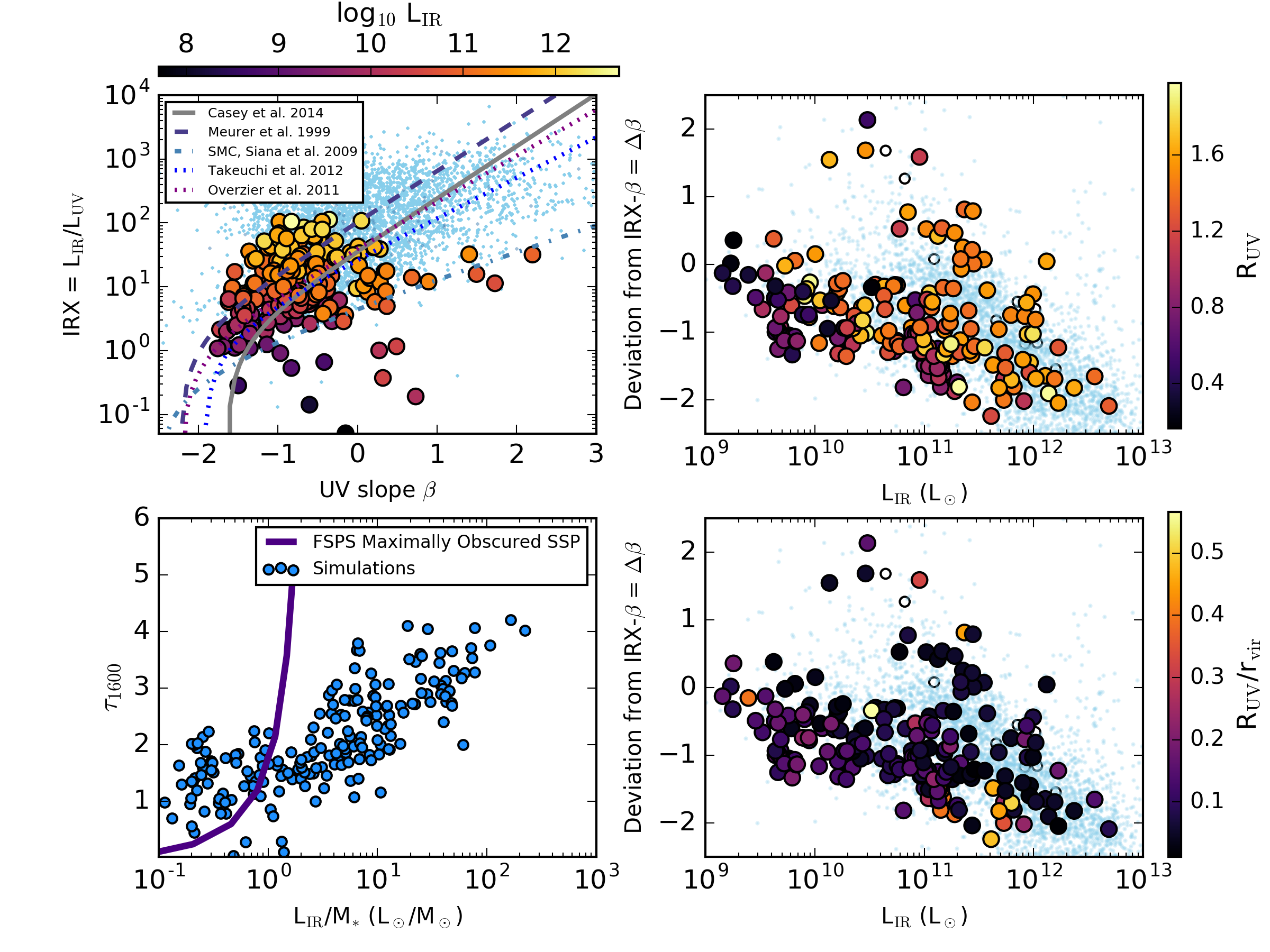}
\caption{ Dusty star forming galaxies with blue UV slopes in the
  \irxb \ plane. The top left panel shows our model galaxies in the
  \irxb \ plane colour-coded by their infrared luminosity.  As is
  evident, galaxies with increasing \lir \ have bluer UV colours as
  compared to the reference relations.  We quantify this in the top
  right panel, where we show the deviation $\Delta \beta$ from the
  \citet{casey14b} relation as a function of \lir.  The blue offset is
  due to two reasons: low UV optical depths due to complex star-dust
  geometries in heavily star-forming galaxies, and (to a lesser
  degree) frosting of UV-bright clumps toward the outskirts of
  galaxies.  The former effect is shown in the bottom left panel,
  where the increase in the UV optical depth with \lir \ (normalised
  by the galaxy stellar mass) is significantly shallower
  than what is expected for complete stellar obscuration (solid line,
  bottom left).  The latter effect is shown by the top and bottom
  right panels, which show the UV radius on an absolute scale, and
  normalised by the galaxy virial radius, as a function of $\Delta
  \beta$.  See text for details.
\label{figure:casey}}
\end{figure*}

Dusty star forming galaxies are the most luminous, heavily
star-forming galaxies in the Universe
\citep[e.g.][]{smail97a,barger98a,hughes98a,hayward11a,hayward13a,narayanan09a,narayanan10b,casey14a}.
A number of works recently have observed that dusty star forming
galaxies at a range of redshifts exhibit particularly blue UV SEDs
given their IRX values as compared to the traditional \irxb
\ relations.  In other words, the UV slopes, $\beta$ are too low for
DSFGs given their infrared excesses.  This was neatly summarised for a
large compiled dataset of $\zsim 0-4$ DSFGs by \citet{casey14b} who
demonstrated that galaxies become {\it bluer} (i.e. larger departures
from the reference \irxb \ relations toward lower $\beta$
slopes) with increasing galaxy infrared luminosity.  Similar effects
have been observed by \citet{penner12a} and \citet{oteo13a}.

In Figure~\ref{figure:casey}, we investigate the origin of blue
infrared bright galaxies.  In the top left panel, we plot our parent
sample, with galaxies colour-coded by their infrared luminosity.
Here, we only plot galaxies at $z>2$ to remain consistent with the
bulk of the observations in these samples.  Immediately it is clear
that more infrared-bright systems present both above and to the left
of the reference \irxb \ relations.  This is further quantified in the
top right panel of Figure~\ref{figure:casey}, where we show the
departure from the reference \irxb \ relations (here, we use the
\citet{casey14b} relation as the reference from which we measure
$\beta$ departures) as a function of galaxy infrared luminosity.  The
light blue points show observations, while the coloured points are our
cosmological zoom simulations.  We ignore the actual colours of the
simulated points for the time being.  Regardless, it is evident that
in both observed galaxies, as well as our models, more
infrared-luminous systems indeed have bluer UV SEDs than the reference
\irxb \ relation.  Moreover, the magnitude of this departure in our
simulations shows good agreement with what is observed.

We first explore the possibility that frosting of UV bright regions
toward the outskirts of galaxies causes the blue offset for high
infrared luminosity galaxies at high-redshift.  Massive galaxies
(i.e. the kind that typically present as DSFGs at high-\z) live in
complex environments, with significant substructure surrounding the
central galaxy \citep[e.g.][]{dave10a,narayanan15a}.  As demonstrated by
\citet{geach16a}, many of these star-forming subhalos are metal poor,
and UV-bright.  In the top right panel of Figure~\ref{figure:casey},
we now highlight the colour-coding, which maps to the UV-luminosity
weighted radius of our model galaxies.  Indeed, this increases by a
factor $\sim 5$ over the dynamic range modeled, with more
infrared-luminous galaxies broadly having larger UV radii.

However, while this likely plays some role in driving the blue offset
of DSFGs from the reference \irxb \ relations, in detail the rise in
UV disk sizes for more infrared luminous galaxies is tempered by the
fact that the galaxies with the highest infrared luminosity also tend
to be the most massive, and hence the largest.  More colloquially,
bigger things are bigger.  We demonstrate this in the bottom right
panel of Figure~\ref{figure:casey}, where we have now normalised the
UV luminosity weighted radii of our model galaxies by their virial
radii.  Indeed, the galaxies with the largest $R_{\rm UV}/R_{\rm vir}$
ratios reside in the bluest, most infrared-luminous locus in $\Delta
\beta$-L$_{\rm IR}$ plane.  However, there are a number of systems
without notably large UV radii.  Hence, frosting plays a role in
driving some DSFGs toward bluer UV colours, though does not
sufficiently paint the entire picture.

Instead, also important are low optical depth sightlines reflective of
a complex star-dust geometry in these systems.  In
Figure~\ref{figure:morphology}, we showed the gas phase morphology for
an arbitrary sample of model galaxies.  The geometries are complex,
and result in significantly lower UV optical depths from decoupled UV
and IR emission sites than what would be expected for a maximally
obscured geometry.  We demonstrate this quantitatively in the
bottom-left panel of Figure~\ref{figure:casey}, where we show the
monochromatic UV ($1600 \ \angstrom$) optical depth as a function of
the infrared luminosity (normalised by galaxy stellar mass) for our
model galaxies.  The rise in $\tau_{1600}$ is shallow with increasing
$L_{\rm IR}/M_*$.  As a comparison, we revisit the \fsps \ numerical
experiment conducted in the top left panel of Figure~\ref{figure:dtm},
where we examined a simple stellar population behind a screen of dust
with increasing optical depth, and plot the optical depth of this
population as a function of \lir/$M_*$ alongside our simulations.
Here, we have slightly modified the numerical experiment from
Figure~\ref{figure:dtm}, in that we enforce that dust dust obscures
all stars, both young and old; this represents maximal obscuration.
There is a dramatic divergence in the $1600 \angstrom$ optical depths
at high specific infrared luminosity between the simplified maximally
obscured stellar population synthesis model, and our hydrodynamic
simulations that exhibit more complex geometries.  At the highest
specific infrared luminosities, the maximal optical depth is
$\tau_{1600} \sim 4-5$, reflective of the large number of sightlines
toward UV sources that are relatively unobscured.

In short, we find that dusty star forming galaxies at high-$z$ exhibit
blue UV slopes due to low optical depth sightlines toward UV-bright
regions. A similar origin is hypothesized by \citet{casey14b}, though
we note that \citet{safarzadeh17a} suggest that frosting (i.e. recent
star formation toward the outskirts of galaxies) may instead dominate
the origin of blue DSFGs.

\subsection{Galaxies in the First Billion Years}
\label{section:highz}

 We now turn our attention to the \irxb \ relation in galaxies at $z
 \ga 5$ \citep[for  recent reviews on UV slopes at these redshifts,
   see][]{stark16a,finkelstein16a}.  The advent of ALMA has allowed for rest-frame
 far infrared detections of galaxies in the early Universe that,
 coupled with a number of assumptions, allows for FIR luminosity
 measurements.  As a result, recent years have seen a number of
 constraints on the \irxb \ relation in low metallicity systems in
 this epoch.  By and large, these galaxies appear to lie below the
 locus of more metal rich galaxies at $z \sim 0-2$ (and even below the
 canonical SMC \irxb \ curve).

As an example, \citet{capak15a} surveyed nine galaxies (between $1-4 \
L_*$ and UV slopes $-1.4 < \beta < -0.7$) at $z \approx 5-6$ in ALMA's
band 7 ($0.8-1.1$ mm observed frame) and found that seven of them lie
below the SMC \irxb \ line (the other two being consistent with the
SMC curve).  Similarly, \citet{bouwens16a} utilised the ALMA
Spectroscopic Survey ($2 \sigma$ detections) in the Hubble Ultra Deep
Field \citep[ASPECS;][]{walter16a,aravena16b,aravena16a,decarli16a} to
constrain the \irxb \ relation in galaxies from $z \sim 2-10$.  Like
the \citet{capak15a} study, \citet{bouwens16a} found many galaxies in
this redshift range to be roughly consistent with the SMC curve (the
more massive systems), or to lie below it.

\begin{figure*}
  \includegraphics[scale=0.9]{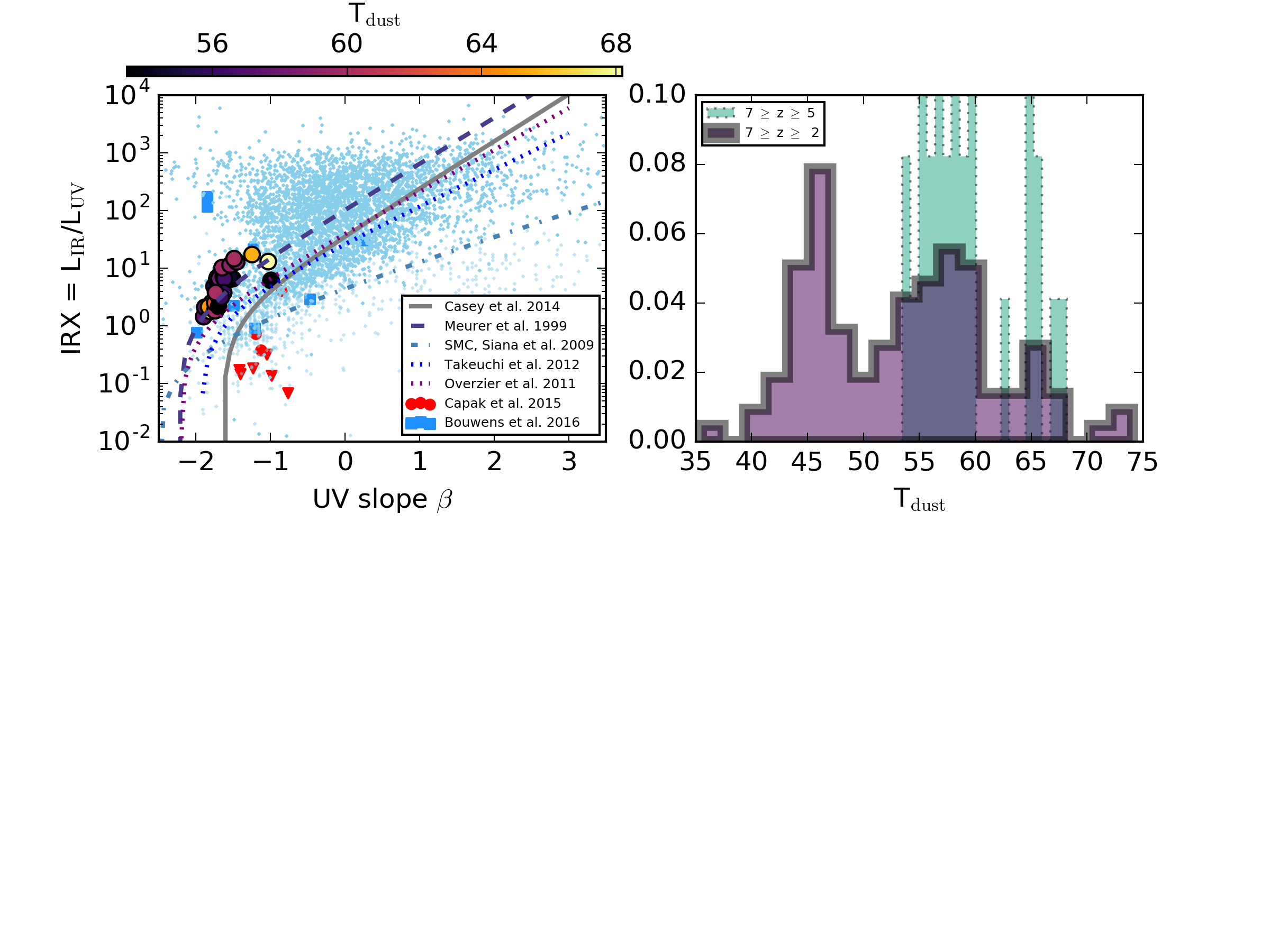}
  \vspace{-5cm}
\caption{\irxb \ Relations for metal poor systems at \z $\gtrsim 5$.
   The left panel shows all model galaxies in our sample between
  $5<z<7$, colour coded by their dust temperature.  Observations of
  systems at these redshifts in particular are noted by the blue
  squares \citep{bouwens16a}, and red circles \citep{capak15a}.
  The downward facing red triangles are upper limits on the
  \citet{capak15a} observations.  While many observations find IRX
  values systematically below the reference relations for these
  galaxies, our models suggest that galaxies of this epoch should lie
  on, or even above the reference relations.  Our models suggest that
  the dust temperatures in these galaxies are quite large ($50-70 $
  K), and observations that assume significantly lower dust
  temperatures (i.e. those representative of $z \sim 2$ galaxies, as
  in the observations presented here) will underestimate \lir, and
  hence IRX.  In the right panel, we show the distribution of $T_{\rm
    dust}$ for all galaxies between $z=2-7$, as well as just those
  between $5<z<7$.  Due to lower dust content and harder radiation
  fields from low metallicity stars, the dust temperatures of the
  highest redshift bin are systematically larger than low redshift
  galaxies. \label{figure:capak_bouwens}}
\end{figure*}

In Figure~\ref{figure:capak_bouwens} we summarise both the detections
(and non-detections) in the aforementioned \citet{capak15a} and
\citet{bouwens16a} studies, as well as show the location of our own
high-redshift model galaxies on the \irxb \ plane.  Here, we show our
model systems between $5 \leq z \leq 7$, and for reference, include
the \citet{casey14b} compilation of $z\sim0-2$ detections.  We
restrict our analysis to models mz0, mz5 and mz10 as these models best
overlap the stellar mass range observed by \citet{capak15a} and
\citet{bouwens16a} ($M_* > 10^9 \msun$ at $z > 5$). As is clear, our
model galaxies all lie on a relatively tight locus near (or above) the
$z\sim 0$ relations, while many of the $z\ga5$ observations have much
lower IRX values.

At face value, the observed lack of dust content in $z\ga5$ galaxies
presents a challenge to our models, which predict that these galaxies
ought to lie on a similar \irxb \ relation as local systems (albeit
toward the low IRX and $\beta$ end).  However, one possibility is that
the observed galaxies have underestimated infrared luminosities (and
hence, IRX values) due to relatively cold assumed dust temperatures.

To elaborate, in deriving the infrared luminosity, \citet{capak15a}
and \citet{bouwens16a} both required an assumed $T_{\rm dust}$ due to
the fact that only one photometric FIR data point was available.
\citet{capak15a} assumed 25 K $\leq T_{\rm dust} \leq$ 45 K, and
\citet{bouwens16a} assumed as a fiducial value, $T_{\rm dust} \sim 35$
K.  Of course if the true dust temperature is warmer than this, then
the derived $L_{\rm IR}$ will increase, and the observed galaxies will
move closer to the local \irxb \ relation.

\begin{figure}
\includegraphics[scale=0.4]{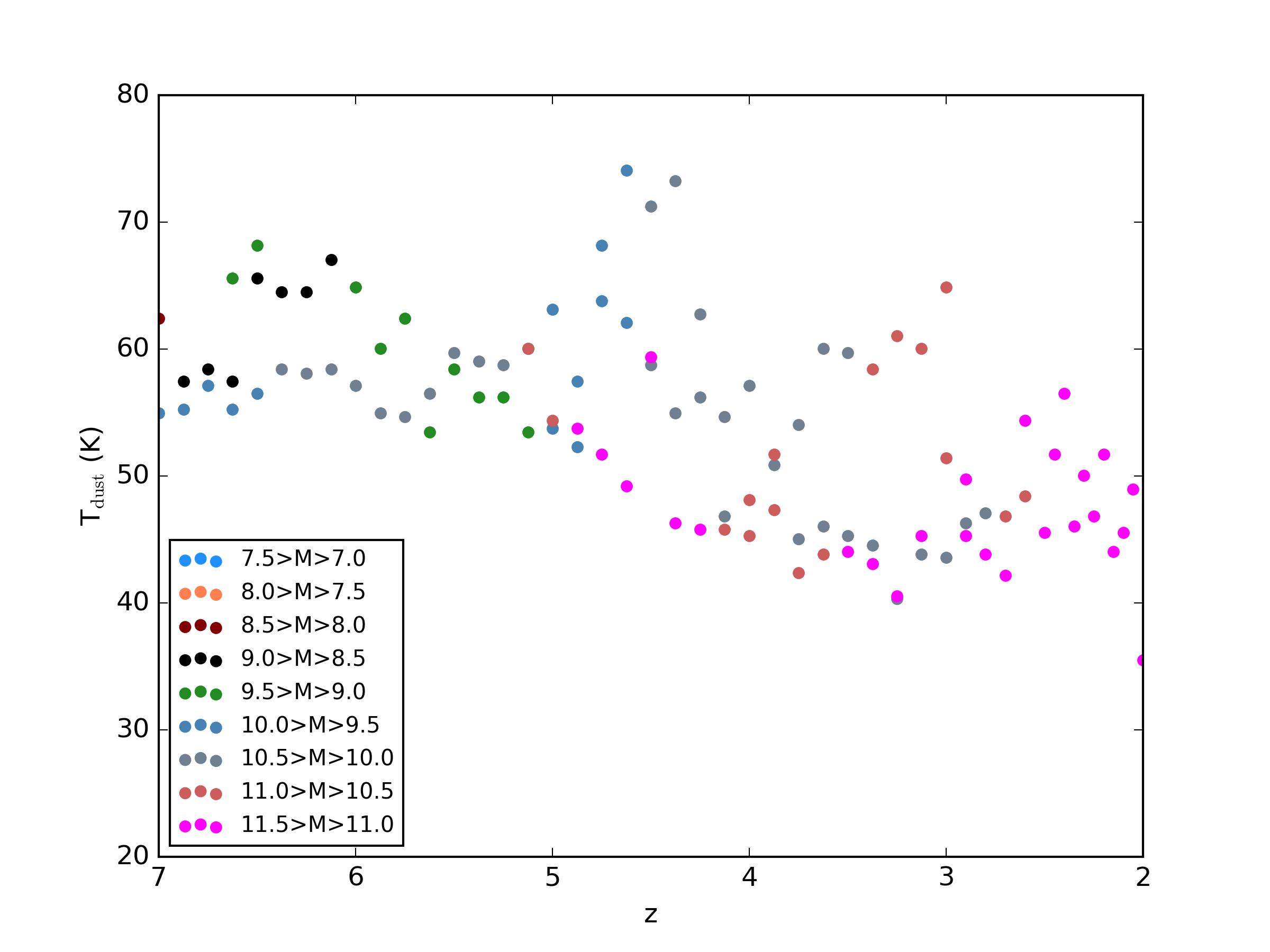}
\caption{Evolution of dust temperature with redshift. We show
  the evolution of $T_{\rm dust}$ for our model galaxies as a function
  of redshift, binned by their stellar mass {\it at that redshift}.
  Decreasing metallicities at higher redshifts cause both lower dust
  contents, as well as harder stellar radiation fields.  The
  combination of these drives up dust temperatures systematically with
  increasing redshift.  As a result, assuming the $T_{\rm dust} \sim
  35-50$ K characteristic of $z \sim 2$ galaxies
  \citep[e.g.][]{magnelli12a} for higher-\z \ sources will cause a
  strong underestimate of the inferred \lir.   \label{figure:tdust}}
\end{figure}

In the right panel of ~\ref{figure:capak_bouwens}, we plot the
distribution of dust temperatures from our model galaxies, showing
both the entire $z > 2$ sample, as well as just the $z > 5$ sample\footnote{To
ensure that we don't miss any cold dust outside our fiducial $25$ kpc
box, we utilise a larger $200$ kpc box for the radiative transfer
calculations for our $T_{\rm dust}$ derivation.}.  The dust temperatures
are derived simply from the peak of the modeled infrared SED.  While
lower redshift ($z \sim 2$) galaxies have dust temperatures comparable
to the assumed observed values of $35-60$ K, at $z \ga 5$, the dust
temperatures rise dramatically, with typical dust temperatures $50 <
T_{\rm dust} < 75$ K.  This owes both to lower dust masses (so UV
photons have lower optical depths), as well as harder radiation fields
in the lower metallicity stars.

On average, dust temperatures drop individual galaxies with 
decreasing redshift.  We show this general case in
\autoref{figure:tdust}, where we show the dust temperature for all of
the galaxies used in Figure~\ref{figure:capak_bouwens}, binned by
their stellar mass at the redshift shown.  A generic trend is that
with increasing redshifts, the average dust temperature of galaxies rises.  Indeed,
\citet{bouwens16a} find that their upper limits would be $\sim 0.4$
dex higher if they had assumed a $T_{\rm dust} = 45-50$ K, which is
still well below the maximum dust temperatures we find at $z \gtrsim 5$ of
$\sim 75$ K.  Even an increase of $\sim 0.4$ dex would be enough to
bring the bulk of the upper limits to values consistent with the SMC curve.

In summary, we therefore find that the typical dust temperature of
galaxies at $z \gtrsim 5$ is substantially larger than what is observed at
lower ($z \sim 2$) redshifts.  When accounting for this increased
$T_{\rm dust}$, inferred infrared luminosities from these systems rise
and bring IRX values of observed galaxies consistent with local \irxb
\ relations.

\section{Discussion}
\label{section:discussion}

\subsection{Secondary (and tertiary) parameters in the Relation}

\begin{figure}
\includegraphics[scale=0.25]{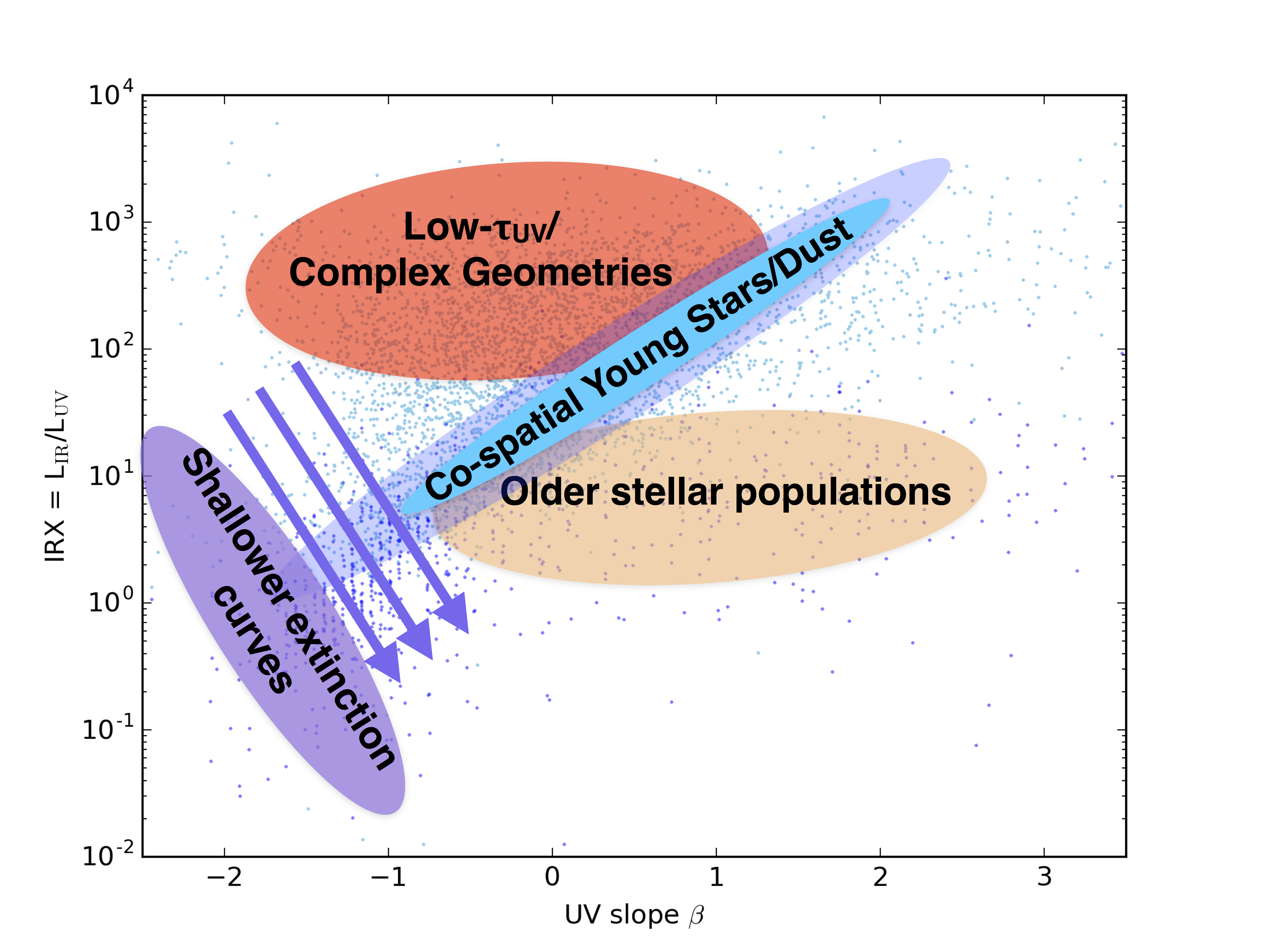}
\caption{Schematic summarising how different physical processes move
  galaxies in the \irxb \ plane.  The origin of these effects are
  described in detail in
  \S~\ref{section:deconstruction}.  \label{figure:cartoon}}
\end{figure}

While the \irxb \ relation appears to hold over a range of physical
conditions, the numerical experiments conducted thus far have
demonstrated a number of scenarios in which galaxies may deviate
strongly from the reference \irxb \ relations.  These deviations have
been observed in the literature for some time.

As a result, a number of authors have investigated the possibility of
``second parameters'' in the \irxb \ relation.  For example,
\citet{kong04a} and \citet{mao14a} have studied the role of star formation history
in driving dispersion in the UV colours in the \irxb
\ relation. \citet{kong04a} defined the 'birthrate parameter',
proportional to ratio of the present to past average SFR.  Galaxies
with lower birthrate parameters move farther from the reference \irxb
\ relations in a perpendicular manner.  At the same time, however,
\citet{johnson07a,johnson13a} find little correlation with the
dispersion in $\beta$ and the star formation history for as
parameterised by the $4000 \angstrom$ break for blue cloud galaxies
(though substantial deviations from the reference \irxb \ relations for
red-sequence galaxies).  Similar results were obtained by
\citet{seibert05a, dale07a} and \citet{cortese08a}.  \citet{grasha13a}
instead find a better relation between the mean stellar age and
deviations from the reference \irxb \ relations.  Indeed, our own
models suggest that the median stellar age is quite relevant in driving red UV slopes, and thereby significant dispersion about the reference \irxb
\ relations (e.g. Figure~\ref{figure:simulations_stellarages}).

Similarly, a number of authors have investigated the dust attenuation
law as a possible driver for dispersion in the \irxb \ relation
\citep[e.g.][]{johnson07b, burgarella05a, boquien09a}.  The dust
attenuation law in galaxies encapsulates both the loss of photons
along the line of sight, as well as scattering back into the line of
sight, and therefore is a proxy for both the extinction law, as well
as radiative transfer and geometric effects.

In our simulations we can identify three major physical
drivers of deviations from the \irxb \ relation.  These are:
\begin{enumerate}

\item Old stellar populations.
  
\item Complex geometries of highly star-forming galaxies (and the resultant decoupled UV and IR emission sites)

\item  Dust extinction curves that deviate substantially from a
  Milky Way-like curve.
  
\end{enumerate}
In Figure~\ref{figure:cartoon}, we summarise the direction of these
deviations schematically.

Because the variation of dust extinction curves with galaxy physical
property is relatively unconstrained \citep[e.g.][]{kriek13a}, we will concentrate on the
impact of older stellar populations and star-ISM geometry in the
remainder of our discussion of additional parameters in the \irxb
\ relation.  While observationally resolving the complexity of the
star-ISM geometry in galaxies (especially at high-\z) is infeasible
for most systems, Figure~\ref{figure:casey} shows that the $\beta$
deviation from the reference \irxb \ relations ($\beta_{\rm ref}$) is
well parameterised by the total infrared luminosity of the galaxy.  It
is apparent that the $\Delta \beta_{\rm ref} \propto \lir$.  Because
the star formation rate is reasonably well represented by the infrared
luminosity in galaxies \citep[within limits; see,
  e.g.][]{younger09a,hayward14a,narayanan15a}, then we can fit $\Delta \beta_{\rm
  ref}$ as a function of the galaxy SFR.  Numerically, we utilise the
best fit relationship from the compendium by \citet{casey14b} as the
reference \irxb \ relation from which we measure deviations in $\Delta \beta_{\rm ref}$:
\begin{equation}
  \label{equation:casey}
  {\rm IRX_{\rm ref}} = 1.68 \times
  \left[10^{0.4[(3.36)+2.04\beta_{\rm ref}]}-1\right]
\end{equation}
and find a relatively good fit with:
\begin{equation}
  \label{equation:deltabeta}
\Delta \beta_{\rm ref} = -0.72 -0.27 \ {\rm log_{10}}\left(\frac{{\rm SFR}}{M_\odot {\rm yr}^{-1}}\right) + \ {\rm exp}\left[1.47\times{\rm log_{10}}\left(\frac{t_{\rm age}}{{\rm Gyr}}\right)\right]
\end{equation}

\begin{figure}
\includegraphics[scale=0.4]{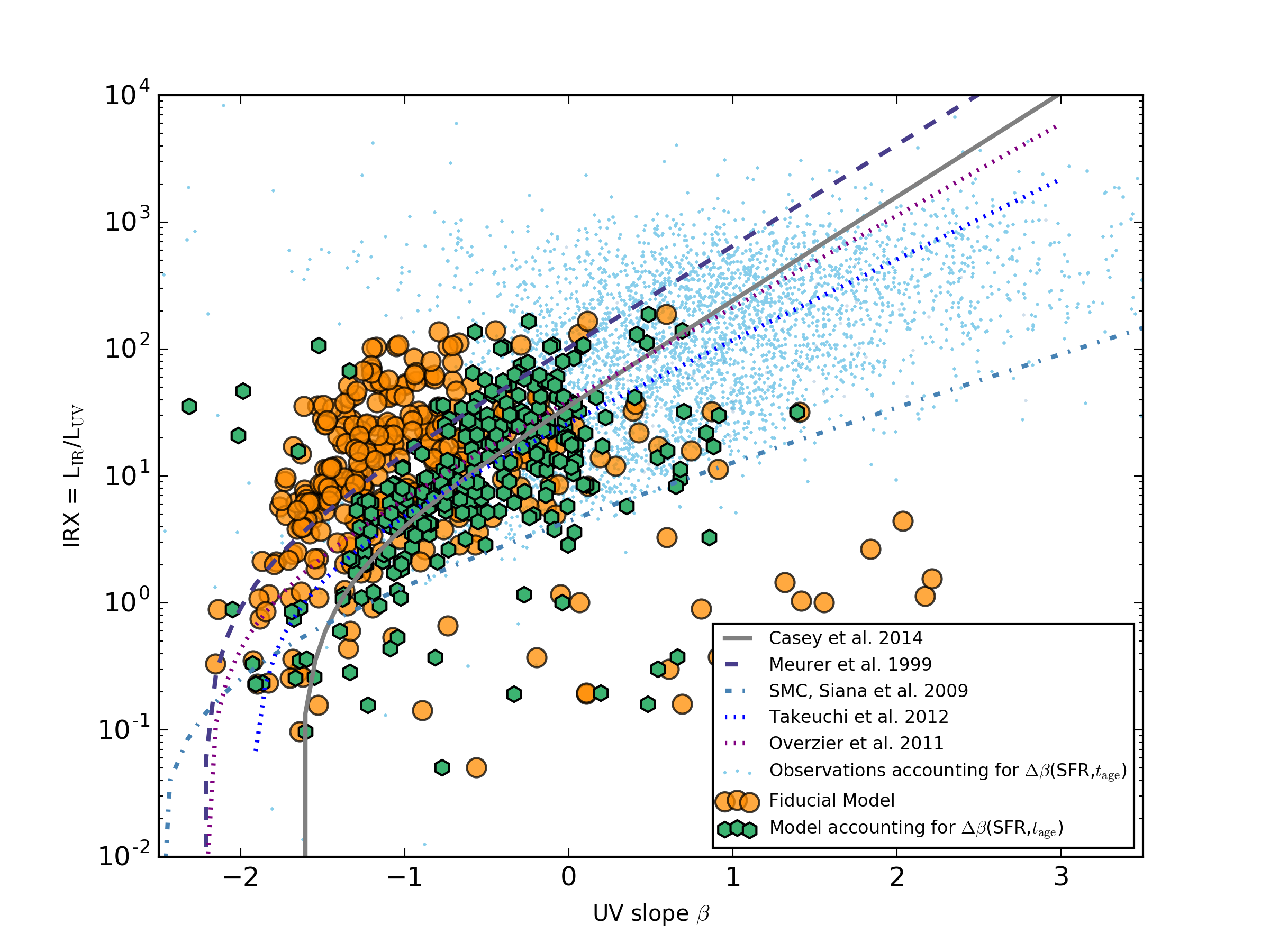}
\caption{\irxb \ relation for our model galaxies and observations
  after correcting for complex star-dust geometries, and older stellar
  populations.  The orange circles are our fiducial model.  The green
  hexagons and faint blue points show our models and observations
  after applying Equation~\ref{equation:deltabeta}, which
  parameterises shifts in the UV slope $\beta$ that occur due to
  decoupled UV and infrared emission (c.f. \S~\ref{section:dsfgs}), as
  well as those that owe to older stellar populations
  (c.f. \S~\ref{section:oldstars}).  The complex geometries are
  reasonably well parameterised by the global SFR.  The shift in UV
  SED slopes represented in Equation~\ref{equation:deltabeta} collapse
  the points with relatively high dispersion toward the reference
  \citet{casey14b} relation given by
  Equation~\ref{equation:casey}. \label{figure:irxb_collapsed}}
\end{figure}

 In Figure~\ref{figure:irxb_collapsed}, we show the \irxb \ relation
 after applying the correction from Equation~\ref{equation:deltabeta}
 (green hexagons).  We find that our model parameterisation for
 $\Delta \beta$ (SFR, $t_{\rm age}$) does a reasonable job at
 collapsing the deviations from our reference \citet{casey14b} \irxb
 \ relation back toward the reference, and substantially reducing the
 dispersion.  We additionally apply the same fit from
 Equation~\ref{equation:deltabeta} to the observational data in
 Figure~\ref{figure:irxb_collapsed}, noting that the dispersion about
 the reference relations in observations as well is significantly
 reduced.  For comparison, in the same figure, we show the fiducial
 \irxb \ relation from our simulations (orange circles), without
 accounting for the blue $\beta$-offset in heavily star forming
 galaxies\footnote{For the observed galaxies, we employ the
   \citet{murphy09a} \lir-SFR calibrations as summarised in
   \citet{kennicutt12a}.  We assume half of the radiation from star
   formation is reprocessed in the infrared.  Because we have no
   information about the stellar ages, we assume $250 $ Myr stellar
   ages for all galaxies.}. By adding the $\Delta \beta$ derived from
 Equation~\ref{equation:deltabeta} to the observed value, one can
 significantly reduce the dispersion and uncertainty inherent in the
 \irxb \ relation.

\subsection{IRX or $\beta$?}
At this point, a natural question is: does IRX or $\beta$ serve as a
better proxy for the UV optical depth?  Indeed, this point has been
investigated by a number of observational studies in recent years
\citep[e.g.][]{cortese08a,kennicutt09a,hao11a}

\begin{figure*}
\includegraphics[scale=0.9]{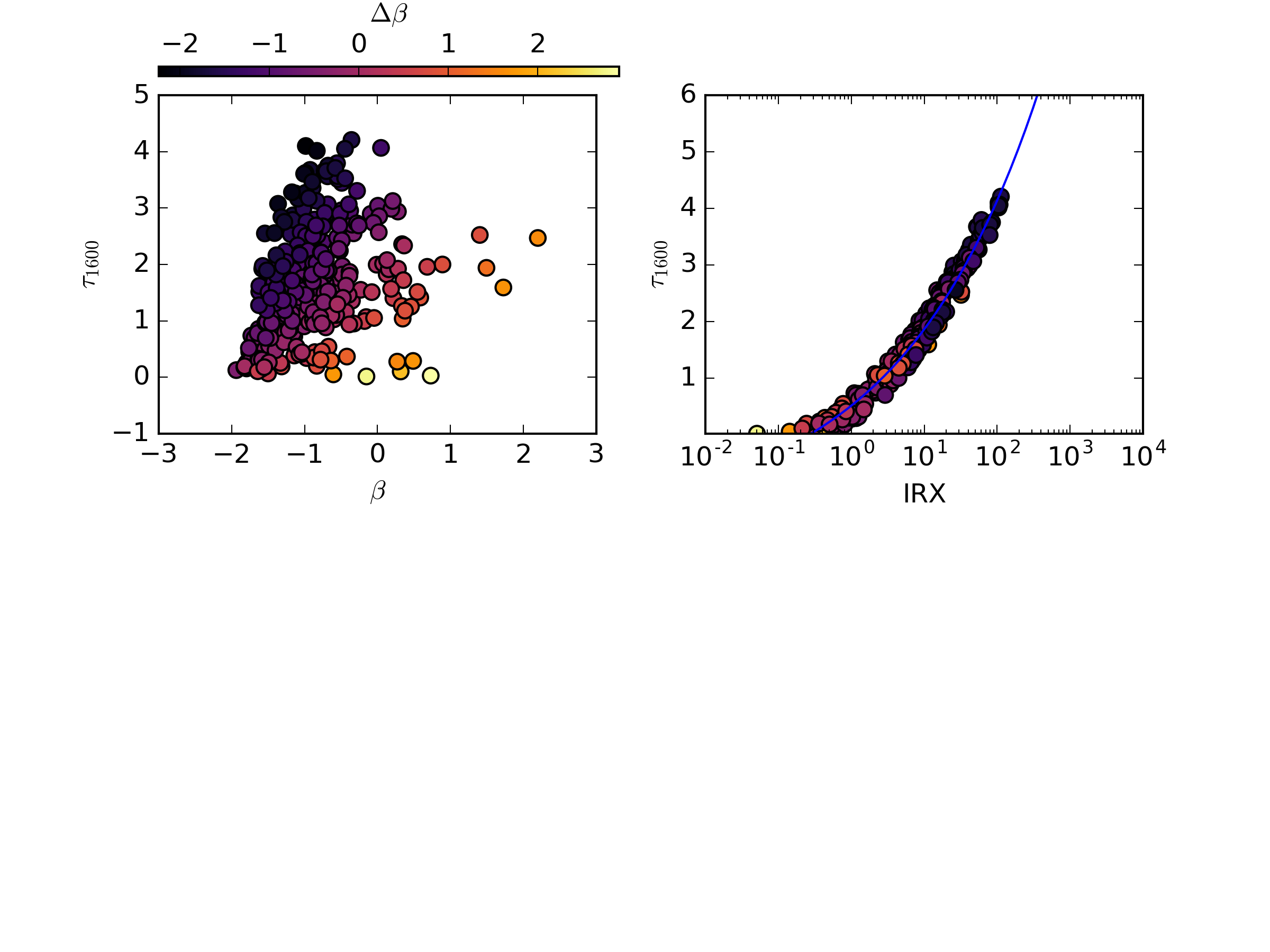}
\vspace{-6cm}
\caption{ UV optical depth ($\tau_{1600}$) vs. $\beta$ (left) and IRX
  (right), colour-coded by the $\beta$ deviation from the
  \citet{casey14b} \irxb \ relation.   While $\tau_{1600}$ shows little correlation with $\beta$, the IRX serves as a good proxy for the UV optical depth, with fit given by Equation~\ref{equation:fuv_optical_depth}.  \label{figure:fuv_optical_depth}}
\end{figure*}

In Figure~\ref{figure:fuv_optical_depth}, we plot $\tau_{1600}$
against both $\beta$ and IRX, colour-coding the points by $\Delta
\beta_{\rm ref}$, where, again we use the \citet{casey14b} derived
relation as our reference relation.  As is clear, $\beta$ generally
does a poor job of representing the monochromatic UV optical depth.  The reason for
this is discussed in \S~\ref{section:dsfgs}.  In short, strongly
varying star-dust geometries complicate the interpretation of $\beta$.

At the same time, IRX serves as a reasonable proxy for $\tau_{1600}$.  We find a relation:
\begin{equation}
  \label{equation:fuv_optical_depth}
\tau_{1600} = 1.47 +   1.98 \times ({\rm IRX}^{0.23})
\end{equation}
well describes the $\tau_{1600}$-IRX relation, and we include this fit
in the right panel (solid blue line) of
Figure~\ref{figure:fuv_optical_depth}.  

\subsection{Comparison to Other Theoretical Models}
A number of works in recent years have attempted to understand the
\irxb \ relation in galaxies from a theoretical standpoint.  By and
large, these methods have employed complementary techniques to our
tool of choice: cosmological hydrodynamic zoom simulations.

In seminal work, \citet{granato00a} coupled the \galform
\ semi-analytic galaxy formation model with dust radiative transfer
calculations to develop a model for infrared-luminous galaxies (with a
particular eye toward local ULIRGs).  Semi-analytic models necessarily
require a simplified star-dust geometry as the structure of galaxies
are evolved in an analytic method.  To wit, \citet{granato00a} assume
that the gas and dust are distributed in exponential disks, with the
dust distributed both in molecular clouds and diffuse (cirrus) ISM.
Stars are allowed to live within molecular clouds within some time
scale $t_{\rm esc}$.  These galaxies are evolved through cosmic time
(with star formation rates/histories determined by the physics
inherent in the analytic prescriptions combined with the modeled halo
growth rates), and processed through \grasil \ dust radiative
transfer.  \grasil, similar to \pd, calculates the stellar SEDs, and
computes the radiative transfer through the dusty ISM.

\citet{granato00a} found that starbursting galaxies within the
\galform \ semi-analytic framework nearly always lie close to the
fiducial \citet{meurer99a} \irxb \ relationship when the stellar
population is young and obscured.  These authors find that the time
scale for which young stars are obscured by their birth clouds,
$t_{\rm esc}$, is one of the more important parameters in dictating
deviations from the reference \irxb \ relations.  These results appear
to be in good agreement with our own model interpretations.  Young
stars that are well hidden by a screen of dust
(c.f. Figure~\ref{figure:dtm}) tend to move along the
reference \irxb \ relations.  When the UV and IR emitting regions
become decoupled (c.f. Figure~\ref{figure:fsps_full_irxb_range}), then
the population exhibits large deviations in the \irxb \ plane.

More recent work by \citet{safarzadeh17a} utilized a set of idealized
hydrodynamic galaxy evolution simulations coupled with \sunrise \ dust
radiative transfer to predict the UV-infrared SED properties of both
idealized disk galaxies, and 1:1 major galaxy mergers.  Here, the
star-dust geometry is determined by the hydrodynamic evolution of the
galaxy/galaxies and has no interaction with cosmologically infalling
gas \citep[this said, it is important to note that substantial radial
  variations in the ISM properties of idealized dics and mergers can
  still occur][]{torrey11a,narayanan09a,narayanan11b}.  Amongst other issues,
these authors investigated both the origin of blue DSFGs in their
simulations.  \citet{safarzadeh17a} show that their models are able to
produce galaxies with blue UV colours, akin to those in the
\citet{casey14b} study, though attribute the origin to 'frosting' of
young stars (i.e. young stars that lie outside the central
dust-obscured nucleus).  This is in contrast to our model, which
demonstrates (Figure~\ref{figure:casey}) that, when normalised by
galaxy virial radius, frosting does not dominate the blue UV colours
of highly star-forming galaxies in a cosmological context.

\citet{safarzadeh17a} additionally investigated the usage of an SMC
dust curve on the location of their idealized model galaxies in the
\irxb \ plane.  Similar to the results found here, these authors
showed that the lack of NUV attenuation drives galaxies toward lower
IRX and redder $\beta$ slopes for the bulk of their star-forming
lives.

\citet{ferrara16a} developed an analytic model for dust growth in
galaxies with the purpose of understanding \citet{capak15a} and
\citet{bouwens16a} $z \ga 5$ galaxies that lie below the reference
\irxb \ relations.  This model derives an equilibrium dust temperature
by balancing the rate of energy absorption with that of emission by
dust grains.  \citet{ferrara16a} find that galaxies with significant
diffuse dust exposed to the UV radiation fields typical of $z \sim
5-6$ Lyman-break galaxies will have warm dust temperatures, and in
effect have true infrared luminosities much larger than what is
inferred when assuming colder dust temperatures typical of lower
redshift galaxies.  This is consistent with our interpretation of
galaxies at $z \ga 5$.  At the same time, these authors point out
that if the clouds in $z \ga 5$ galaxies are denser, then the majority
of the dust will be shielded from the UV radiation field, and at lower
temperatures.  These galaxies will have low IRX values, consistent
with the \citet{capak15a} and \citet{bouwens16a} interpretations.

\subsection{Possible Tensions with Observations}

We now turn to areas where our models may conflict with observational
results.  \citet{reddy10a} examined a sample of $90$ Lyman break
galaxies at $z \sim 2$ that had both infrared (Spitzer) and UV
measurements.  With the aid of SED modeling, these authors determined
that young galaxies ($t_{\rm age} < 100 $) Myr have systematically lower
IRX and redder $\beta$ values than the older systems ($t_{\rm age} >
100 $ Myr in their sample.  At face value, this is in conflict with
 our model results presented in
Figures~\ref{figure:simulations_stellarages} and
~\ref{figure:fsps_stellarages}.

One possibility is that the young galaxies in the \citet{reddy10a}
sample have an intrinsic SMC-type extinction curve.  When these
authors calculate IRX and $\beta$ for their young galaxies assuming
an SMC-like dust curve, the observed young galaxies all lie relatively
close to the SMC curve, which could then be accommodated by our model
(c.f. Figure~\ref{figure:smc}).  While it is unknown what drives
variations in extinction and attenuation laws in galaxies, certainly a
broad range of attenuation laws are observed in galaxies at low and
high-redshift.  Indeed, \citet{kriek13a} find that galaxies with
larger specific star formation rates tend to have shallower dust
attenuation curves.  If the younger systems in the \citet{reddy10a}
sample have shallower attenuation curves (i.e. comparable to an SMC
curve), then the tension between our model and the \citet{reddy10a}
observations may be reduced.  This said, \citet{reddy06a} and
\cite{reddy10a} suggest that attenuation laws are steeper for younger
UV-selected galaxies.

\section{Summary}
\label{section:summary}
We have developed a theoretical model for the origin of and variations
in the \irxb \ dust attenuation relation in galaxies.  To do this, we
have combined cosmological zoom hydrodynamic galaxy formation
simulations with stellar population synthesis models and 3D dust
radiative transfer calculations.  Additionally, we compare these to a
variety of stellar population synthesis models.  Our main results
follow:

\begin{enumerate}
  \item Galaxies with relatively young stellar populations, a Milky
    Way-like extinction curve, and relatively cospatial IR and UV
    emitting regions tend to lie on or near the standard
    \citet{meurer99a} or \citet{casey14b} local relations.  As the
    dust content in galaxies increase, galaxies move along these
    relations.

    \item Substantial variations can be present from these reference
      relations.  These generally owe their origin to the following
      effects:

      \begin{itemize}

      \item Older stellar populations tend to lie below the canonical
        \irxb \ relations, due to changes in the UV-optical SEDs of
        evolved stars.

        \item Complex star-dust geometries tend to drive galaxies
          above the canonical relations, due to low optical-depth
          sightlines that cause galaxies to have bluer $\beta$ values
          than those that have more cospatial IR and UV emitting
          regions.  These complex geometries are common in
          high-redshift heavily star-forming galaxies.

          \item Galaxies with SMC-like extinction curves, or those
            with a Milky Way-like extinction curve (but without a
            $2175 \angstrom$ UV bump) lie below the canonical \irxb
            \ relations.

      \end{itemize}

    \item We have used these results to understand the origin of
      deviations from the \irxb \ relation in both high-redshift dusty
      star forming galaxies, as well as those detected at $z > 5$.
      The former class of galaxies have relatively blue UV SEDs due to
      complex star-dust geometries, and low optical depth sightlines
      toward UV bright regions.  The latter class of galaxies tend to
      lie on or even above the canonical \irxb \ relation.  This said,
      their dust temperatures are quite warm ($50-70 $ K), so IRX
      inferences based on a single long-wavelength photometric point
      are subject to systematically underestimating the \lir \ if
      assuming a lower $T_{\rm dust}$.

    \item We use the results from these simulations to derive two
      fitting relations to maximize the utility of the \irxb
      \ relation in galaxies.
      \begin{itemize}
        \item We derive a fitting relation
          (Equation~\ref{equation:deltabeta}) between $\Delta
          \beta_{\rm ref}$, $t_{\rm age}$ and the SFR (i.e. the
          deviation in $\beta$ from the reference relations due to
          complex geometries [characterized by the galaxy SFR] and the
          stellar ages).  The usage of this correction factor reduces
          the dispersion inherent in the \irxb \ relation.

          \item We derive a fit between the UV optical depth
            ($\tau_{1600}$), and IRX (which serves as a good proxy for
            the UV optical depth);
            Equation~\ref{equation:fuv_optical_depth}.
      \end{itemize}
      
\end{enumerate}

In general, our models suggest that no single \irxb \ relation exists
for galaxies, and that some caution must be exercised when correcting
for dust obscuration using this method.  We find that it is possible
to correct for some of these deviations, though a reasonable amount of
scatter in the relationship is still present.

\section*{Acknowledgements}
The authors thank Caitlin Casey, Alex Pope, Naveen Reddy, Nick
Scoville, Brian Siana and Dan Stark for helpful conversations.  We are
additionally appreciative to Rychard Bouwens, Caitlin Casey, Brian
Siana, and Fabian Walter for providing the results from their
observational studies in easily digestable digital format.
D.N. thanks the organizers of the ``Physical Characteristics of Normal
Galaxies at \z$>2$'' conference in Leiden, where many of the ideas for
this paper were generated.  The simulations published here were run on
the University of Florida HiPerGator supercomputing facility, and the
authors are grateful to the University of Florida Research Computing
for providing computational resources and support that have
contributed to the research results reported in this publication.
Partial support for DN was provided by NSF AST-1009452, AST-1445357,
AST-1724864, NASA HST AR-13906.001 from the Space Telescope Science
Institute, which is operated by the Association of University for
Research in Astronomy, Incorporated, under NASA Contract NAS5-26555,
and a Cottrell College Science Award, awarded by the Research
Corporation for Science Advancement.  JEG acknowledges the Royal
Society for a University Research Fellowship.

\end{document}